\documentclass[10pt,aps,prb,nobibnotes,twocolumn,groupedaddress]{revtex4-1}

\usepackage[utf8]{inputenc}
\usepackage{placeins}
\usepackage{tabularx}
\usepackage{amsmath}
\usepackage{amsfonts}
\usepackage{nicefrac} 
\usepackage[colorlinks,citecolor=blue,urlcolor=blue,bookmarks=false,hypertexnames=true]{hyperref}
\usepackage{amssymb}
\usepackage{graphicx}
\usepackage{float}
\usepackage{threeparttable}
\usepackage{subcaption}
\usepackage[dvipsnames]{xcolor}
\usepackage[shortlabels]{enumitem}
\usepackage{dcolumn} 
\usepackage{braket}
\newcolumntype{d}[1]{D{.}{.}{#1}}
\newcommand{\inlinecite}[1]{\citenum{#1}}

\begin{document}
\def\ie{{\it i.e.\/}}
\def\eg{{\it e.g.\/}}
\def\etal{{\it et al.\/}}
\def\cm{{cm$^{-1}$}}
\def\kJmol{kJ$\,$mol}
\def\Eh{$E_\mathrm{h}$}
\def\mEh{m$E_\mathrm{h}$}
\def\uEh{$\mu E_\mathrm{h}$}
\def\a0{$a_0$}
\def\bea{\begin{eqnarray}}
\def\eea{\end{eqnarray}}

\title{Surface adsorption at the thermodynamic limit using periodic DLPNO-MP2 theory: \\ A study of CO on MgO at dilute and dense coverages}
\author{Andrew Zhu}
\affiliation{University of Oxford, South Parks Road, Oxford, OX1 3QZ, UK}
\author{Poramas Komonvasee}
\affiliation{University of Oxford, South Parks Road, Oxford, OX1 3QZ, UK}
\author{Arman Nejad}
\affiliation{University of Oxford, South Parks Road, Oxford, OX1 3QZ, UK}
\author{David P. Tew}
\affiliation{University of Oxford, South Parks Road, Oxford, OX1 3QZ, UK}

\date{\today}

\begin{abstract}
We apply periodic domain-based local pair natural orbital second-order M{\o}ller--Plesset perturbation theory (DLPNO-MP2) to probe the adsorption energy of CO on MgO(001), the consensus toy model system for surface adsorption. A number of robust correlated wavefunction methods now achieve excellent agreement with experiment for the adsorption of a single CO molecule onto the MgO surface. However, studies probing denser coverage ratios are scarce because of the increased computational expense and the larger configuration space to optimize. We leverage the computational efficiency of periodic DLPNO-MP2 to perform simulations beyond a single unit cell. By using large supercells, we highlight the importance of accurately representing the thermodynamic limit of the surface, and demonstrate in turn that different coverage ratios can be consistently probed. In the dilute regime, we show that adsorption energies obtained from periodic DLPNO-MP2 agree with existing benchmarks. We then obtain adsorption energies at increasing densities approaching full monolayer coverage. Our results show a reduction in binding strength at full coverage, agreeing with experimental observations, which is explained by the increasing lateral repulsions between the COs. This study demonstrates the efficacy of periodic DLPNO-MP2 for probing increasingly sophisticated adsorption systems at the thermodynamic limit.
\end{abstract}
\maketitle

\section{Introduction}
The interaction of a molecule as it adsorbs onto a surface is a key chemical phenomenon, where accurate modeling provides benefits to a number of important applications. In particular, computational evaluation of the adsorption energy gives mechanistic insight to reactions for heterogeneous catalysis, gas storage and surface lubrication, among others. \cite{christensen_molecular_2008, somorjai_active_1977, chen_computational_2021, ertl_elementary_1990, schauermann_nanoparticles_2013, vogt_concept_2022, morris_gas_2008, jahanmir_adsorption_1986}  Electronic structure studies of adsorption systems have predominantly employed density functional theory (DFT), due to its computational efficiency. However, accurate modeling of the dispersion interactions involved is difficult unless semi-empirical corrections are incorporated into the density functional treatment.\cite{grimme_semiempirical_2006, grimme_consistent_2010, grimme_density_2011, tkatchenko_accurate_2009, dion_van_2004, ugliengo_are_2002} Notwithstanding these dispersion-corrected functionals (DFT-D), higher accuracy quantum chemistry methods, which inherently capture dispersive effects, are needed to provide reliable interaction energies.

The adsorption of a single carbon monoxide (CO) molecule onto a pristine magnesium oxide (MgO(001)) surface has become the consensus toy model system to probe surface adsorption interactions. Computational schemes can be roughly divided into finite-cluster methods, or approaches employing periodic boundary conditions.
Recently, a number of studies, employing high accuracy correlated wavefunction methods, including second-order M{\o}ller--Plesset perturbation theory (MP2) and coupled cluster theory with singles, doubles and perturbative triples excitations (CCSD(T)), have obtained excellent agreement for the adsorption energy of CO on MgO at the dilute coverage limit\cite{boese_accurate_2013,alessio_chemically_2019,shi_many-body_2023,ye_adsorption_2024}.
Alessio et al.\cite{alessio_chemically_2019} and Boese and Sauer\cite{boese_accurate_2013} both employ hybrid MP2: DFT-D embedded cluster calculations, incorporating single point CCSD(T) calculations to obtain the final adsorption energy estimate, with values of $-21.2 \pm 0.5 \;\mathrm{kJ\; mol^{-1}}$ and $-21.0 \pm 1.0 \;\mathrm{kJ\; mol^{-1}}$, respectively. Alessio et al. also employ periodic local MP2 to demonstrate agreement with their embedded cluster MP2 estimates. Shi et al.\cite{shi_many-body_2023} used CCSD(T) calculations within their embedded cluster SKZCAM approach, obtaining an adsorption energy of $-19.2 \pm 1.0 \;\mathrm{kJ \; mol^{-1}}$, which they show agreement with canonical periodic CCSD(T) and diffusion Monte Carlo. Finally, Ye and Berkelbach\cite{ye_adsorption_2024} calculate adsorption energies using periodic local natural orbital (LNO) schemes at MP2 and CCSD(T) levels, calculating a value of $-20.0 \pm 0.5 \; \mathrm{kJ \; mol^{-1}}$. Considering that the touted standard of `chemical accuracy' is within $4.2 \;\mathrm{kJ\; mol^{-1}}$, the overall consensus of these methods is remarkable, considering the differences in the approaches taken and the need to converge out errors in all the method specific wavefunction and basis related parameters, as well as finite-size effects. 

Experimental enthalpies for CO on MgO(001) adsorption have been determined using temperature-programmed desorption (TPD) techniques, of which we highlight the work from Dohn\'alek et al.,\cite{dohnalek_physisorption_2001} and Wichtendahl et al.\cite{wichtendahl_thermodesorption_1999} Both studies produce TPD spectra at varying CO coverage ratios ($\Theta$), reporting desorption values at minimal densities of around $\frac{1}{4}$ monolayer coverage, which provide the closest estimate to the adsorption energy at the dilute limit. Using these experimental references, previous computational works have conducted analyses to compute comparable adsorption energy values, by accounting for thermal and zero-point energy contributions, the PV term, and the pre-exponential factor used in the Redhead equation\cite{redhead_thermal_1962}. Boese and Sauer obtained a TPD-derived energy of $(-20.6 \pm 2.4)$\,kJ\,mol$^{-1}$, whilst Shi et al. obtained an experimental estimate of $(-19.2 \pm 1.0)$\,kJ\,mol$^{-1}$. With both experiment-derived and theoretical studies from a number of different works agreeing to well within chemical accuracy, the adsorption energy of CO on MgO at the dilute regime appears to have reached a robust consensus, providing a rigorous benchmark.

There are, however, still unanswered questions for simulating adsorption processes. In real chemical conditions, adsorption reactions rarely feature a single adsorbate molecule, and consideration for the optimal ratio of surface site availability is a key concern for heterogeneous catalysis. For example, experimental literature for CO adsorption onto an MgO surface reports varying reactivities associated with dilute to dense coverage regimes.\cite{shigeishi_chemisorption_1976, tait_n_2006, dohnalek_physisorption_2001, wichtendahl_thermodesorption_1999, schmid_analysis_2023} Whilst an agreed benchmark for the adsorption energy of a single CO molecule on a pristine MgO surface has now been established, simulations incorporating multiple CO adsorption sites at denser monolayer coverages are still scarce. These systems are inherently more difficult to simulate, given the expanded configuration space to now optimize over, and the need to incorporate lateral interactions between the adsorbate monomers. For these larger adsorption unit cells or fragments, simulating the in-principle infinite extent of the surface, or the thermodynamic bulk limit, becomes increasingly important in order to remove finite size errors or edge effects. All in all, the increased computational demand required to simulate larger surface systems poses a steep challenge for current correlated wavefunction methods.

Recently, we have outlined the theory and implementation of two complementary methods for periodic domain-based local pair natural orbital (DLPNO) second-order M{\o}ller--Plesset perturbation theory\cite{nejad_dlpno-mp2_2025-1, zhu_dlpno-mp2_2025}. These schemes enable a vast compression of the virtual space, enabling calculations of large supercells to accurately extrapolate to the thermodynamic limit. In this contribution, we use periodic DLPNO-MP2 to probe CO on MgO(001) adsorption with large supercells beyond a single unit cell. We focus upon ensuring our calculations are converged with respect to the thermodynamic limit of the surface, which then enables us to robustly probe the adsorption energy at different CO coverage densities.

First, we outline our computational methodology employing periodic DLPNO-MP2 to model surface adsorption systems. Then, in Section \ref{sec:dilute-mgoco}, we verify the validity of DLPNO-MP2 for surface interactions by evaluating the adsorption energy towards the infinitely dilute regime, which can be benchmarked rigorously against the aforementioned schemes. Having established agreement, we then use periodic DLPNO-MP2 to probe CO coverage ratios approaching the fully filled monolayer ($\Theta =$1) limit, in Section \ref{sec:dense-mgoco}. This is achieved through leveraging the computational efficiency of periodic DLPNO-MP2, enabling simulation of multiple adsorption sites within a supercell, whilst ensuring the thermodynamic limit of the surface is probed. In doing so, we have conducted some of the largest supercell calculations to probe surface adsorption using correlated wavefunction schemes.

\section{Methods}
\subsection{Periodic DLPNO-MP2}

DLPNO theory\cite{neese_efficient_2009, riplinger_efficient_2013} is an established method within molecular quantum chemistry to reduce the computational cost of correlated wavefunction schemes, achieving near-linear scaling of computational effort with system size with only modest loss in accuracy by replacing integrals and excitation amplitudes with low-rank approximations that exploit the inherent locality of electron correlation in insulators. Our two works, BvK-DLPNO-MP2 and Megacell-DLPNO-MP2, represent the first full adaptations of DLPNO theory to periodic systems.

Our `BvK' scheme employs Born--von K{\'a}rm{\'a}n (BvK) boundary conditions\cite{nejad_dlpno-mp2_2025-1}, whilst the `Megacell' method retains rigorous translational symmetry but exchanges the lattice summation in the BvK integrals with explicit sums over direct-space interactions\cite{zhu_dlpno-mp2_2025}. Both implementations converge to the same values at the thermodynamic limit and also show excellent agreement with canonical benchmarks. The degree of compression through the PNOs is controlled through a single variable, the occupation number threshold, $\mathcal{T}_{\mathrm{PNO}}$, where the error incurred due to discarding virtuals is proportional to $\sqrt{{\mathcal T}_\text{PNO}}$. We demonstrate numerically that the pilot scheme of Megacell-DLPNO-MP2 is particularly computationally efficient, with near-linear scaling with respect to supercell size. Our aim in this contribution is to use Megacell-DLPNO-MP2 to reveal further insight into interactions of the MgO + CO adsorption system. Previous computational studies of the surface adsorption energy of CO on MgO have incorporated higher accuracy correlated wavefunction methods, notably CCSD(T). Whilst our study employs only MP2, we note that previous work demonstrates that MP2 is almost as accurate as CCSD(T) for these interactions\cite{ye_adsorption_2024,ugliengo_are_2002,herschend_influence_2006}, suggesting significant value in a periodic MP2 method that can efficiently simulate the thermodynamic bulk of the surface layer in an adsorption reaction.

\begin{figure}[htbp]
    \includegraphics[width=\linewidth]{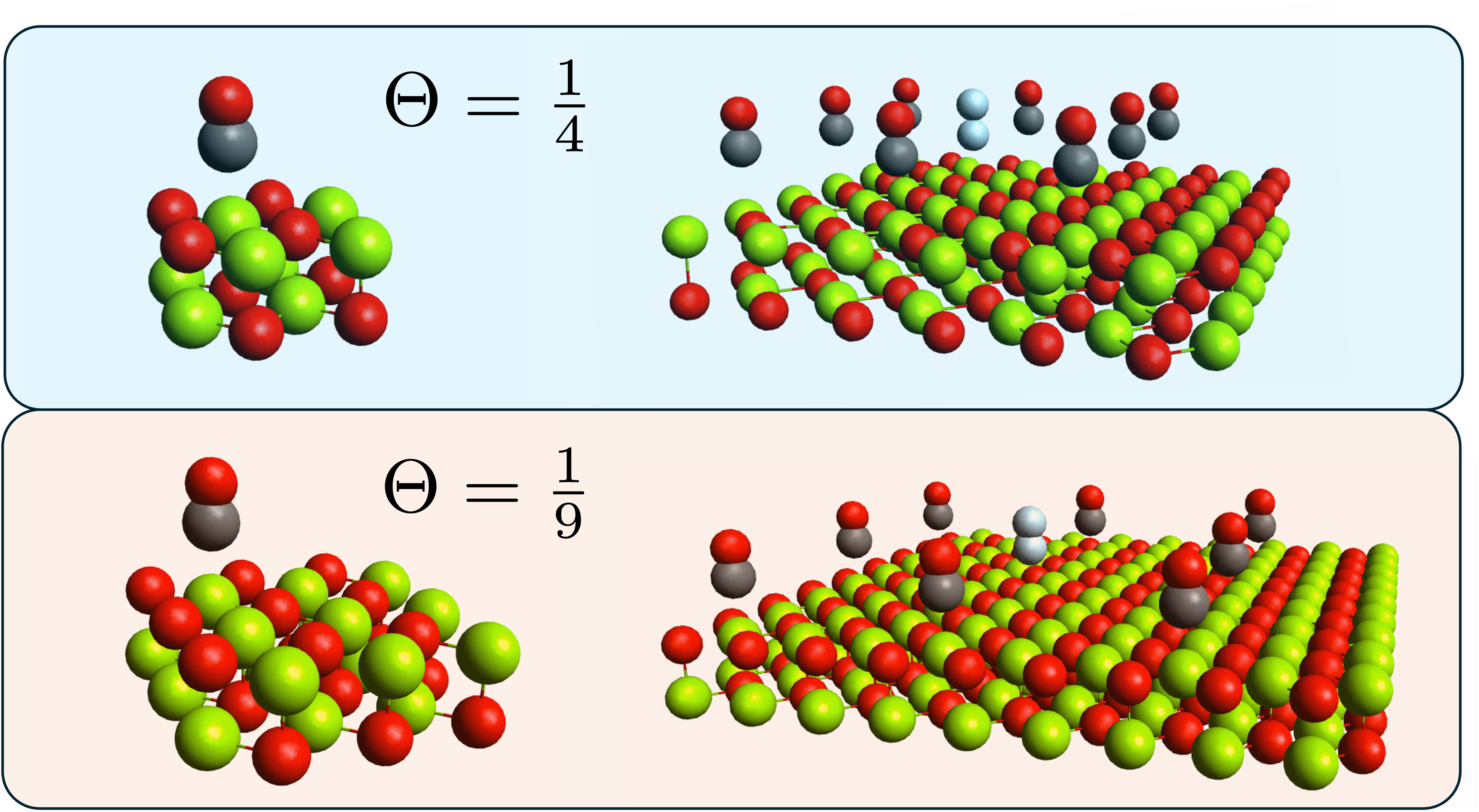}
    \caption{Unit cell structures for CO adsorption on an MgO surface corresponding to coverage ratios of $\Theta=\frac{1}{4}$ (top) and $\Theta = \frac{1}{9}$ (bottom). The unit cells are depicted on the left, the $3{\times}3$ supercells on the right where the central cell CO molecule is highlighted.}
    \label{fig:tdl}
\end{figure}

The efficacy of periodic DLPNO-MP2 enables multiple unit cells, each featuring a CO adsorption site, to be simulated. Figure \ref{fig:tdl} presents unit cells corresponding to coverage ratios of $\Theta=\frac{1}{4}$ and $\frac{1}{9}$, respectively, as well as 3 $\times$ 3 supercells of each system, where the CO molecule in the reference (central) unit cells are highlighted in white. By simulating more than a single unit cell (Gamma point), our calculations using Megacell-DLPNO-MP2 ensure that the adsorbed molecule in the reference cell has an interaction length scale with the surface that can extend beyond the boundaries of the unit cell. This ensures the correct thermodynamic limit of the bulk surface is always being probed. In turn, accurate evaluation of the different coverage ratios can then be controlled by the size of the surface slab and the number of adsorbates within the unit cell. In this contribution we predominantly employ $3\times3$ supercells, using $5\times5$ supercells occasionally to verify the thermodynamic limit convergence of our calculations. We believe this work is the first instance where a periodic post-HF scheme simulates multiple adsorption sites in a periodic supercell calculation, in contrast to previous studies which employ only single unit cells or Gamma point calculations.

\subsection{Evaluation of Adsorption Energy}

\begin{figure}[htbp]
    \includegraphics[width=\linewidth]{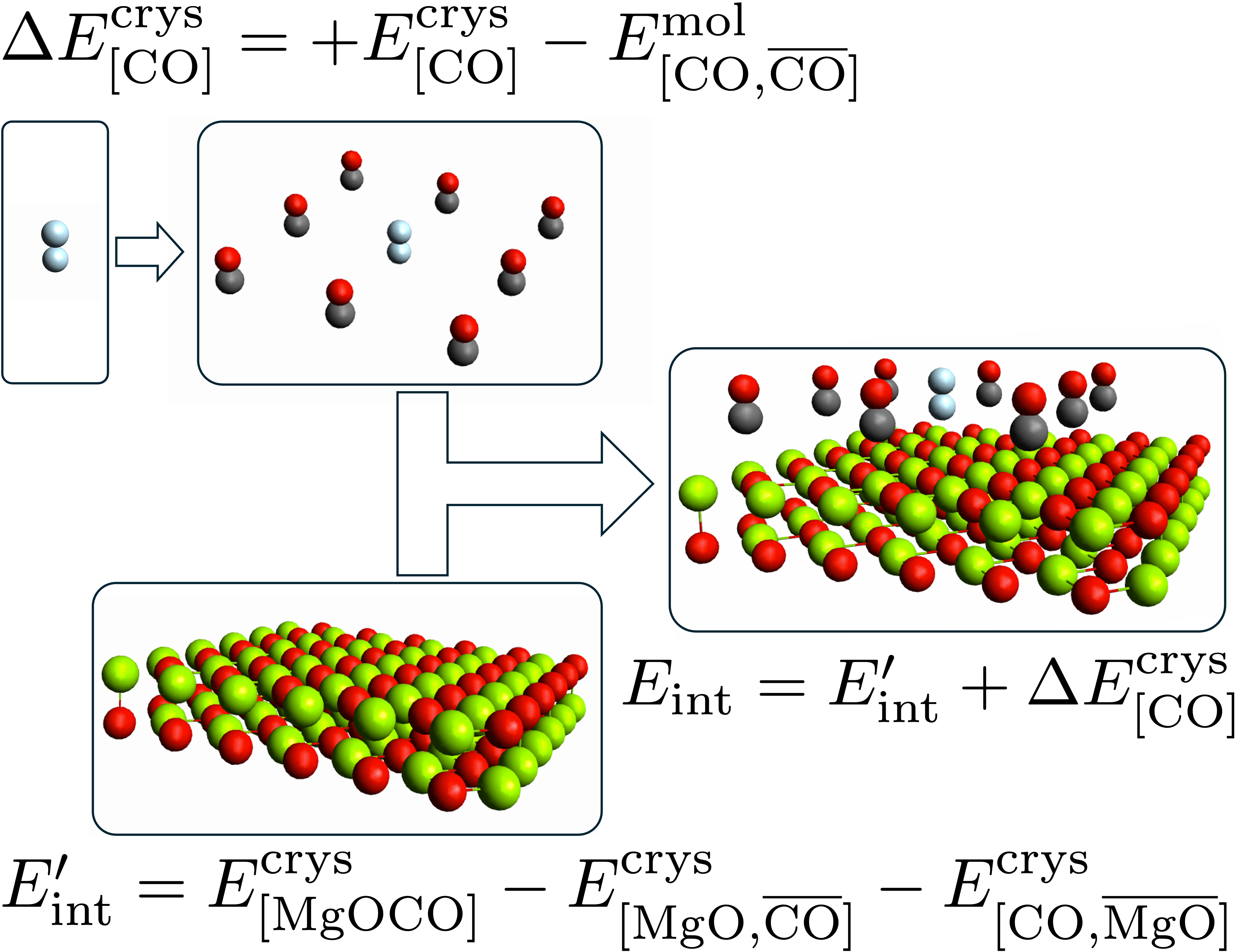}
    \caption{The contributions to the adsorption energy of CO on the MgO surface, for any given coverage ratio. The lateral interactions of the gas phase CO molecules must be added to $E_{\mathrm{int}}$ to properly distinguish adsorption energies at different coverage ratios.}
    \label{fig:eintcoh}
\end{figure}

The adsorption energy is defined as the energy difference, per unit cell, between the adsorbed CO molecule on the MgO surface, and the energies of the individual non-interacting components,
\begin{equation} \label{eq:Adsenergy}
    E_{\mathrm{ads}} =  E_{[\mathrm{MgOCO}]}^\text{crys,eq}-E_{[\mathrm{MgO}]}^\text{crys,eq}-E_{[\mathrm{CO}]}^\text{mol,eq},
\end{equation}
where $[\mathrm{MgO}]$ is the pristine MgO surface and $[\mathrm{CO}]$ is the isolated gas-phase CO molecule and each energy is evaluated at the fully relaxed equilibrium structure. In line with previous work\cite{alessio_chemically_2019,shi_many-body_2023,ye_adsorption_2024}, we decompose $E_{\mathrm{ads}}$ into
\begin{equation} 
    E_{\mathrm{ads}} =  E_{\mathrm{int}} + \Delta_\text{geom}
\end{equation}
where $E_{\mathrm{int}}$ is the interaction energy of CO and MgO, computed using MgO and CO geometries frozen at those of the CO + MgO system, and $\Delta_\text{geom}$ is the energy change upon relaxing to the geometries of pristine MgO and gas-phase CO.  $E_{\mathrm{int}}$ is computed with counter-poise correction to reduce basis set superposition error (BSSE)
\begin{align}
    \label{eq:Eint}
    E_{\mathrm{int}} &= E^\prime_{\mathrm{int}} + \Delta E_\mathrm{[CO]}^\text{crys}, \\
    E^\prime_{\mathrm{int}} &= 
    E^\text{crys}_{[\mathrm{MgOCO}]}-E^\text{crys}_{[\mathrm{MgO,\overline{CO}}]}-E^\text{crys}_{[\mathrm{CO,\overline{MgO}}]} , \\
     \Delta E_\mathrm{[CO]}^\text{crys} &=
    + E_\mathrm{[CO]}^\text{crys} - E_\mathrm{[CO,\overline{CO}]}^\text{mol} .
\end{align}
Here a system ${[X,\overline{Y}]}$ denotes the $X$ subsystem with ghost functions representing $\overline{Y}$.  The second term $\Delta E_\mathrm{[CO]}^\text{crys}$ accounts for the lateral interactions between CO molecules and is the difference between a periodic calculation of the CO lattice and a molecular calculation using ghost functions at the lattice sites. Many previous works approximate $E_{\mathrm{int}} \approx E^\prime_{\mathrm{int}}$, neglecting the second term, which is indeed insignificant at dilute CO coverage. We are interested in examining the dense coverage regime, where this term becomes significant. Figure \ref{fig:eintcoh} is a schematic visualizing the contributions within $E_{\mathrm{int}}$. The final adsorption energy in this paper is thus given by, 
\begin{equation} \label{eq:Adsenergy}
    E_{\mathrm{ads}} =  E^\prime_{\mathrm{int}}+\Delta E_\mathrm{[CO]}^\text{crys} + \Delta_{\mathrm{geom}}.
\end{equation}
 Ye and Berkelbach\cite{ye_adsorption_2024} and Shi et al.\cite{shi_many-body_2023} both compute $\Delta_{\mathrm{geom}}$ using informed choices of DFT functionals, incorporating dispersion corrections, reaching a similar agreement of approximately $1.0 \; \mathrm{kJ \; mol^{-1}}$ for dilute coverages.

\subsection{Computational Details}

Within our study, Megacell-DLPNO-MP2 is employed to evaluate all quantities contributing to $E_{\mathrm{int}}$, apart from $E^{\mathrm{mol}}_{[\mathrm{CO,\overline{CO}}]}$. This calculation is aperiodic, featuring only a single CO molecule surrounded by ghost functions, which we compute using the existing molecular DLPNO-MP2 implementations\cite{schmitz_scaling_2013,tew_principal_2019} in the \verb;pnoccsd; module, within the TURBOMOLE program. The HF orbitals for this system are obtained from the \verb;ridft; module.

Megacell-DLPNO-MP2\cite{zhu_dlpno-mp2_2025} has been implemented in a developmental version of the TURBOMOLE\cite{balasubramani_turbomole_2020} program, within the \verb;pnoccsd; module\cite{schmitz_scaling_2013,schmitz_explicitly_2014,schmitz_perturbative_2016,tew_principal_2019}. Periodic LCAO-based HF calculations using $k$-point sampling have recently become available on developmental branches of TURBOMOLE, in the \verb;riper; module\cite{irmler_robust_2018,lazarski_density_2015,lazarski_density_2016,burow_linear_2011,burow_resolution_2009,muller_real-time_2020}, the output of which provides the HF Bloch functions and band energies required for MP2. The RI-J approximation is employed and we adopt the Monkhorst-Pack grid for our $k$-point grid\cite{monkhorst_special_1976}. We use our recently developed Wannier function (WF) localization procedure to prepare localized occupied orbitals\cite{zhu_wannier_2024}. As described in Ref.\inlinecite{zhu_dlpno-mp2_2025}, Megacell-DLPNO-MP2 embeds a supercell correlation treatment within a larger megacell in order to ensure all correlated WFs are sufficiently decayed. Periodic HF calculations are performed on a $k$-grid spanning the size of the megacell, which is related to the supercell size by $k_{\mathrm{mega}}=2k_{\mathrm{super}}-1$ in each cartesian dimension.

Previous work from Shi et al.\cite{shi_many-body_2023} highlights the underestimation of $E_{\mathrm{int}}$ if correlation from the 2s2p orbitals on magnesium is neglected. We therefore only freeze the 1s core shells for C, O and Mg in our DLPNO-MP2 treatment. In our investigation for the dilute regime, we directly use the optimized unit cell geometries reported by Ye and Berkelbach\cite{ye_adsorption_2024}, which we describe in greater detail in Section \ref{sec:dilute-mgoco}. A pristine surface of MgO is employed for the dense coverage calculations. All unit cell geometries are reported in the SI. 

All calculations were run on a single node (Intel(R) Xeon(R) Gold 6248R CPU) with a maximum RAM limit of 386\,GB and 1.8\,TB disk, with OMP parallelization of up to 48 threads. No single calculation in this contribution exceeded three days of wall times, highlighting the efficiency of the Megacell-DLPNO-MP2 approach. Our largest calculations featured supercells containing just under 30000 orbital basis functions, representing the largest calculations undertaken by Megacell-DLPNO-MP2 thus far.

\subsection{Basis Set}
The importance of using sufficiently expansive basis sets to capture the adsorption interactions has been highlighted extensively in previous works. In the initial phase of testing, we encountered issues converging the periodic HF calculations using all-electron cc-pVTZ orbital basis sets\cite{dunning_gaussian_1989,woon_gaussian_1993}, which we attributed to linear dependency issues arising from the diffuse basis functions centered on Mg, leading to divergent exchange contributions. HF convergence using the all electron pob-TZVP\cite{peintinger_consistent_2013} and pob-TZVP-rev2\cite{vilela_oliveira_bssecorrection_2019,laun_bsse-corrected_2022}  basis sets presented no issues. However, diffuse and additional polarization functions are vital for accurately capturing the dispersion interactions in adsorption systems, but are absent from these reduced basis sets. Our solution was to employ the cc-pVXZ orbital basis sets for C and O, and to use modified pob-XZVP-rev2 basis sets for Mg (X=D,T). These modified basis sets contained additional valence polarization and diffuse functions from the cc-pVXZ basis sets, as well as core polarisation functions from the cc-pwCVXZ basis sets.  We thus performed two sets of adsorption calculations, using approximate `DZ' and `TZ' quality basis sets, enabling an estimate of the complete basis set limit values. The basis sets employed are tabulated in the SI. 

For the RI-J approximation within the periodic HF calculations, the cc-pVTZ auxiliary basis sets\cite{weigend_accurate_2006}  were employed for carbon and oxygen atoms, whilst the def2-TZVP auxiliary basis sets\cite{weigend_accurate_2006}  were used for magnesium, for all calculations. The same auxiliary basis sets were also employed for the density-fitting treatment within the MP2 calculations, apart from for the ghost atoms, which did not feature any auxiliary basis functions.

\subsection{Canonical and Basis Set Extrapolations}

In order to provide meaningful comparison with other theoretical and experimental adsorption energies, our results are extrapolated to the respective canonical and basis set limits. For each individual quantity contributing to $E_{\mathrm{int}}$, DLPNO-MP2 calculations are performed at two occupation number thresholds ($\mathcal{T_{\mathrm{PNO}}}=10^{-7},10^{-8}$), with both the `DZ' and `TZ' basis sets. For each basis set, a square root extrapolation to estimate the complete PNO space (CPS) limit\cite{sorathia_basis_2020,sorathia_improved_2024} is first performed,
\begin{equation} 
    E_{\mathrm{corr,CPS}} =  \frac{\mathcal{T}_2^{1/2}E(\mathcal{T}_1)-\mathcal{T}_1^{1/2}E(\mathcal{T}_2)}{\mathcal{T}_2^{1/2}-\mathcal{T}_1^{1/2}}.
\end{equation}
Here, $E(\mathcal{T}_{1})$ and $E(\mathcal{T}_{2})$ are DLPNO-MP2 correlation energies obtained at two occupation number thresholds, where $\mathcal{T}_{1}>\mathcal{T}_{2}$. With these CPS estimates, the correlation energy estimate at the complete basis set (CBS) limit is then obtained through Helgaker's two-point extrapolation\cite{helgaker_priori_2004,kutzelnigg_principle-quantum-number_2008},
\begin{equation} \label{eq:helgaker}
    E_{\mathrm{corr,CBS}} =  \frac{X^3 E(X)-Y^3 E(Y)}{X^3-Y^3},
\end{equation}
where $X$ and $Y=X-1$ refer to the cardinality of the basis sets used. We then compute $E_{\mathrm{int}}$ from Eq(\ref{eq:Eint}). The dominant source of uncertainty arises from the basis set extrapolation. We also compute a $E_{\mathrm{int}}$ value using the $E_{\mathrm{corr,TZ}}$ correlation energies, and define the uncertainty as half the difference between this value and the basis set extrapolated $E_{\mathrm{int}}$ energy. Finally, the total DLPNO-MP2 interaction energy is evaluated by summing the extrapolated correlation contribution to the HF energies, which are computed in the `TZ' basis.

\begin{table}[htbp]
\centering
    \caption{Hartree--Fock and correlation energy contributions to $E'_{\mathrm{int}}$ and $E_{\mathrm{int}}$, comparing supercell sizes of $3\times3$ and $5\times5$. Energies are given in $\mathrm{kJ \; mol^{-1}}$. Calculations employed the $2\times2$ surface slab unit cell, using the modified `TZ' basis set. Total MP2 interaction energies are given, as the sum of the HF and CPS correlation energy components.}

    \label{tab:tdl}
    \begin{ruledtabular}
    \begin{tabular}{|l|cc|cc|}
      & \multicolumn{2}{c|}{$E'_{\mathrm{int}}$} & \multicolumn{2}{c|}{$E_{\mathrm{int}}$} \\
    $k_{\mathrm{super}}$ & $3\times3$ & $5\times5$ & $3\times3$ & $5\times5$ \\
    \hline
    HF  & 2.13 & 2.13 & 2.47 & 2.47\\
    $E_{\mathrm{corr}}(\mathcal{T}_{\mathrm{PNO}}=10^{-7})$ & -17.71 & -17.79 & -18.42 & -18.51 \\
    $E_{\mathrm{corr}}(\mathcal{T}_{\mathrm{PNO}}=10^{-8})$ & -18.26 & -18.38 & -18.97 & -19.11\\
    $E_{\mathrm{corr}}$ (CPS) & -18.52 & -18.66 & -19.23 & -19.39 \\
    \hline
    MP2 (CPS) & -16.38 & -16.53  & -16.76 & -16.91 \\
    \end{tabular}
    \end{ruledtabular}
\end{table}

\section{Towards the dilute coverage limit}
\label{sec:dilute-mgoco}

In this section, we study the adsorption of a single CO molecule on the pristine MgO surface, modeling the surface at the thermodynamic limit, and probing the adsorption energy towards the infinitely dilute limit. We follow similar protocols to Ye and Berkelbach\cite{ye_adsorption_2024} and Shi et al.\cite{shi_many-body_2023}, who model the MgO(001) surface using a two-layer slab, which both studies conclude is sufficient to converge the adsorption energy with respect to the layer depth. The CO molecule is adsorbed in a perpendicular fashion to the surface, with the C end pointing towards the five-fold coordinated Mg site, with a Mg--C interaction distance set to 2.460\,\AA, to enable direct comparison with the aforementioned two works. 

We employ three types of unit cell to probe the dilute regime: a $2\cdot2$, $3\cdot3$ and $4\cdot4$ surface slab of 2-layer MgO, corresponding to surface coverages of $\Theta=\frac{1}{4}$, $\frac{1}{9}$. and $\frac{1}{16}$. $2\cdot2$ and $3\cdot3$ unit cells are presented in Figure \ref{fig:tdl}. Our chosen geometries are taken directly from the optimized equilibrium structures reported by Ye and Berkelbach\cite{ye_adsorption_2024}, which they obtain through geometry optimization with the Perdew–Burke–Ernzerhof\cite{perdew_generalized_1996} (PBE) functional with D3 dispersion treatment\cite{grimme_consistent_2010}, fixing the bottom layer surface slab. They report a single geometry relaxation energy from all surface slab optimizations, $\Delta_{\mathrm{geom}}$, of $1.1 \; \mathrm{kJ \; mol^{-1}}$, and we thus adopt the same value. The crucial difference with our study is that our supercell periodic approach can independently examine different coverage ratios, whilst also converging the thermodynamic limit of the surface through increasing the supercell size.

\subsection{Dilute Regime Results}
\label{sec:diluteregimeres}


We first verify the thermodynamic limit convergence of the adsorption calculations. Table~\ref{tab:tdl} presents the HF and correlation energy contributions to the interaction energy for the $2\cdot2$ surface slab unit cell at `TZ' quality, for supercell sizes of $3\times3$ and $5\times5$. The correlation energies are computed at two $\mathcal{T_{\mathrm{PNO}}}$ thresholds, which are then used to estimate the CPS limit using a square root extrapolation\cite{sorathia_basis_2020,sorathia_improved_2024}. MP2 energies, summing the HF and CPS values, are also given. The full interaction energies, $E_{\mathrm{int}}$, and the contribution ignoring CO lateral interactions, $E'_{\mathrm{int}}$, are both presented.

Examining the HF energies, we first note that the $3\times3$ and $5\times5$ supercells correspond to HF calculations spanning the megacell, which uses $k$-grids of size $5\times5$ and $9\times9$, respectively. Agreement, to a precision of $0.01 \mathrm{kJ \; mol^{-1}}$, is achieved, indicating thermodynamic limit convergence. For correlation energies, including the extrapolated CPS limits, we note small deviations between the two supercell sizes, of magnitudes around $0.1 \mathrm{kJ \; mol^{-1}}$. The close agreement of these energies, especially when compared to the magnitude of uncertainty when converging to the basis set limit, which we subsequently discuss, allows us to conclude that both the thermodynamic limit of the surface slab, and the canonical limit have been sufficiently converged. Comparing $E_{\mathrm{int}}$ and $E'_{\mathrm{int}}$, we see that lateral interactions between CO molecules have a very small contribution to the overall correlation energy in this dilute regime, of less than $1 \mathrm{kJ \; mol^{-1}}$. Computing energies for the $3\cdot3$ and $4\cdot4$ surface slab unit cells at $5\times5$ supercell sizes is currently difficult due to the severe memory demands of these very large supercells, but given that the $2\cdot2$ unit cell represents the densest coverage ratio, we expect $3\times3$ supercells to also be sufficient for representing the thermodynamic limit for these unit cells.

\begin{figure}[htbp]
    \centering
    \includegraphics[scale=0.50]{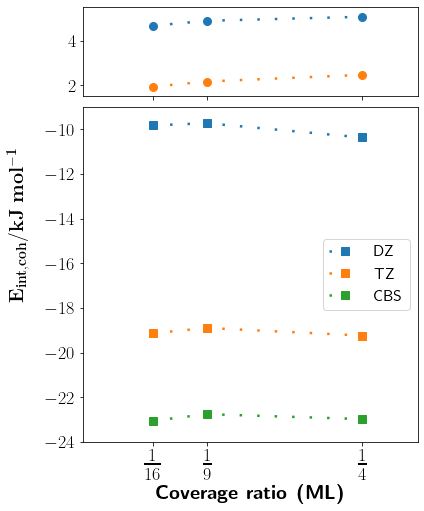}
    \caption{Hartree--Fock (Top) and MP2 correlation energies (Bottom) for $E_{\mathrm{int,coh}}$, at three dilute coverage ratios. Results using the modified `DZ' and `TZ' basis sets are shown, and complete basis set estimate is provided for the MP2 correlation energies.}\label{fig:Eint-coh-res}
\end{figure}

Figure \ref{fig:Eint-coh-res} presents the $E_{\mathrm{int}}$ energies at increasing coverage ratios, obtained from the $4\cdot4$, $3\cdot3$ and $2\cdot2$ surface slab unit cells, respectively. The top panel presents the HF energies, whilst the bottom plots the MP2 correlation energy contributions. For both cases, the cohesive energies for different coverage ratios are evaluated using the `DZ' and `TZ' basis sets. A $3\times3$ supercell was employed in all cases. For the MP2 correlation energies, an inverse cube extrapolation is performed to estimate the basis set limit value, using Eq.(\ref{eq:helgaker}).

We first highlight that the cohesive energies are largely similar in magnitude across the different coverage ratios, indicating that all three coverage ratios capture the dilute coverage regime. Whilst a slight increase in MP2 correlation energy is observed from $\frac{1}{4}$ to $\frac{1}{9}$ coverage ratios, this deviation is on a similar order of magnitude to the uncertainties incurred from the CPS and thermodynamic limit extrapolations, and thus not significant. The HF cohesive energies show a very slight decrease towards further dilute coverages. The observation that all three ratios are within the same dilute regime agrees with the experimentally derived adsorption energies reported by Dohn{\'a}lek et al.\cite{dohnalek_physisorption_2001}. They report, in Figure 4 of Ref.\inlinecite{dohnalek_physisorption_2001}, an adsorption energy at $\frac{1}{4}$ coverage that is within $1 \mathrm{kJ \; mol^{-1}}$ of their extrapolated adsorption energy at zero coverage. 

For our method, we stress that the use of the $3\times3$ supercell size is crucial in obtaining the correct dependence with coverage ratio. This is because surface-adsorbate interactions extending beyond the length scale of the unit cell are explicitly considered in the supercell calculation, meaning the thermodynamic limit of the surface is more accurately represented. In contrast, Ye and Berkelbach\cite{ye_adsorption_2024} perform single point (unit cell) calculations, and report a significant decrease in correlation energy from the $2\cdot2$ to $4\cdot4$ unit cells with periodic MP2. We attribute this behavior to the growing surface slab size, which substantially increases the magnitude of the surface interaction if only one unit cell is considered. In the case of Ref.\inlinecite{ye_adsorption_2024}, where the motivation is to extrapolate to the infinitely dilute regime, this approach is completely valid. However, since we intend to discern differences between coverage ratios with Megacell-DLPNO-MP2, any interaction energy dependence on the surface slab size of the unit cell must be factored out. Calculations beyond a single unit cell are thus necessary in order to remove the finite size error associated with the surface.

Figure \ref{fig:Eint-coh-res} shows that extrapolation of the `DZ' and `TZ' quality basis sets to a basis set limit estimate gives a significant decrease in MP2 correlation cohesive energies compared to the `TZ' values, on the order of $4\;\mathrm{kJ \; mol^{-1}}$, across all coverage ratios. This source of error is the largest component of uncertainty within our scheme, and is admittedly significantly larger than previous computational schemes, which we record in Table \ref{tab:bench}. As mentioned, we are currently prevented from performing calculations using cc-pVTZ or higher quality basis sets due to our periodic HF. Work is currently underway to address these issues, which will enable calculations with QZ quality basis sets, to obtain narrower error bars, as demonstrated by Ye and Berkelbach\cite{ye_adsorption_2024}.

\begin{table}[htbp]
    \centering
    \caption{Comparison of adsorption energies of CO on the MgO(001) surface using Megacell-DLPNO-MP2 and recent results from the literature at various surface coverage ratios ($\Theta$). `cl' and `pbc' refer to cluster and periodic wavefunction schemes, respectively. Experimental data from TPD spectroscopy\cite{wichtendahl_thermodesorption_1999,dohnalek_physisorption_2001} have been converted to adsorption energies in two different analyses\cite{boese_accurate_2013,shi_many-body_2023}.}
    \label{tab:bench}
    \begin{tabular}{|c|c|c|}
    \hline
    $\Theta$ & $E_{\mathrm{ads}}\;(\mathrm{kJ \; mol^{-1}})$ &  Methodology\\
    \hline
    $1/4$       & -19.4 $\pm$ 1.9 & Megacell-DLPNO-MP2 \\ 
    $1/9$      & -19.5 $\pm$ 1.9 & Megacell-DLPNO-MP2 \\
    $1/16$      & -20.0 $\pm$ 2.0 & Megacell-DLPNO-MP2 \\
    &&\\[-1em]
    \hline
    dilute limit & $-21.2 \pm 0.5$ & cl MP2:DFT-D+$\Delta$CC\cite{alessio_chemically_2019}\\ 
    dilute limit & $-21.6 \pm 0.3$ & pbc MP2\cite{alessio_chemically_2019}\\
    $1/8$        & $-21.0 \pm 1.0$ & cl MP2:DFT-D+$\Delta$CC\cite{boese_accurate_2013}\\
    dilute limit & $-18.8 \pm 0.3$ & pbc MP2\cite{ye_adsorption_2024}\\ 
    dilute limit & $-20.0 \pm 0.5$ & pbc CCSD(T)\cite{ye_adsorption_2024}\\ 
    dilute limit & $-18.5 \pm 0.5$ & cl MP2\cite{shi_many-body_2023}\\ 
    dilute limit & $-19.2 \pm 0.6$ & cl CCSD(T)\cite{shi_many-body_2023} \\ 
    \hline
    dilute limit          & $-20.6 \pm 2.4$ & TPD\cite{boese_accurate_2013} \\ 
    dilute limit       & $-19.2 \pm 1.0$ & TPD\cite{shi_many-body_2023} \\ 
    \hline
    \end{tabular}
\end{table}

The final adsorption energies obtained via Megacell-DLPNO-MP2 are presented in Table \ref{tab:bench}, compared to previous periodic or cluster based schemes, as well as experimental TPD results. Our values are obtained by summing the HF energy for the `TZ' basis with the CBS extrapolated correlation energy. The three coverage ratios probed all fall within the dilute coverage regime, and we note the general excellent agreement of our values, within $2\;\mathrm{kJ \; mol^{-1}}$ of all other values presented in the Table. This is not surprising, given previous evidence of the good agreement of periodic and cluster MP2 calculations with higher level coupled cluster results. Clearly, our larger error bars, obtained from halving the difference between the `TZ' and CBS extrapolated correlation energies, are a point of concern, stemming from our limitations in basis set. Despite this, the overall agreement of the scheme gives promise for Megacell-DLPNO-MP2 as a robust scheme to probe different adsorption coverages.

\section{Probing dense coverages}
\label{sec:dense-mgoco}

Whilst numerous wavefunction studies probing the adsorption energy of CO on MgO at the dilute coverage limit achieve excellent agreement with experimental values, considerably less work has been done to simulate the adsorption energy dependence at higher coverage ratios. From the perspective of simulations, the difficulty is now the vastly increased degrees of freedom that now need to be analyzed. Due to the lateral interactions between CO molecules, which are significant at higher coverage densities, differing (i.e. non-perpendicular) CO orientations and configurations need to be considered, in order to obtain the lowest energy state for a given coverage ratio. Whilst some experiments give details of highly ordered c(4x2) phases\cite{audibert_co_1992,panella_co2_1994} at ratios close to full monolayer coverage, relatively little consensus has been established for favored phases between $\frac{1}{4}$ to $\frac{3}{4}$ coverages, providing no benchmark systems for computational approaches. Dohn\'alek et al.\cite{dohnalek_physisorption_2001} provide an experimentally-derived adsorption energy plot as a function of CO coverage, which confirms a decrease in the magnitude of the adsorption energy towards full monolayer coverage, due to increasing CO lateral repulsions.

Cluster calculations are inherently disadvantaged, since modeling multiple CO adsorption sites to capture the CO lateral interactions involves building a much larger fragment. Recent work by Shi et al.\cite{shi_accurate_2025} provides a solution by evaluating the cohesive energy of the lattice of COs in the absence of the surface at CCSD(T) accuracy, and incorporating a correction for the effect of the MgO surface using DFT. However, lateral interactions under the effect of the surface are never directly explored to correlated wavefunction accuracy. Periodic approaches are more naturally suited to modeling denser coverage ratios, since the adsorbed layer increasingly resemble periodic models. However, large supercells, beyond single unit cell calculations, are required to correctly capture the full extent of the periodic lateral interactions. Furthermore, larger unit cells are needed if differing CO orientations or configurations are to be probed. Computational studies are scarce, but we highlight the work by Minot et al.\cite{minot_theory_1996}, using periodic Hartree--Fock, which report a combination of perpendicular and bent CO geometries at $\Theta=\frac{3}{4}$, presenting two different arrangements which are energetically comparable. To our knowledge, no correlated wavefunction studies, targeting denser coverage ratios, exist, and even studies with DFT are scarce.

In this section, we exploit the efficiency of Megacell-DLPNO-MP2 to simulate supercells of dense CO coverage. We choose not to explore the full parameter space involving non-perpendicular CO geometries or differing configurations, which would require a substantially more involved systematic study. Instead, we use Megacell-DLPNO-MP2 to simulate unit cells featuring perpendicular COs at increasingly dense coverage ratios, approaching $\Theta=1$. Although not directly comparable to experimental TPD studies, we still expect to capture the effect of the lateral repulsions between COs with our simplified scheme, and observe the decreasing magnitude of $E_{\mathrm{ads}}$, towards $\Theta=1$. 
\begin{figure}[htbp]
    \includegraphics[width=\linewidth]{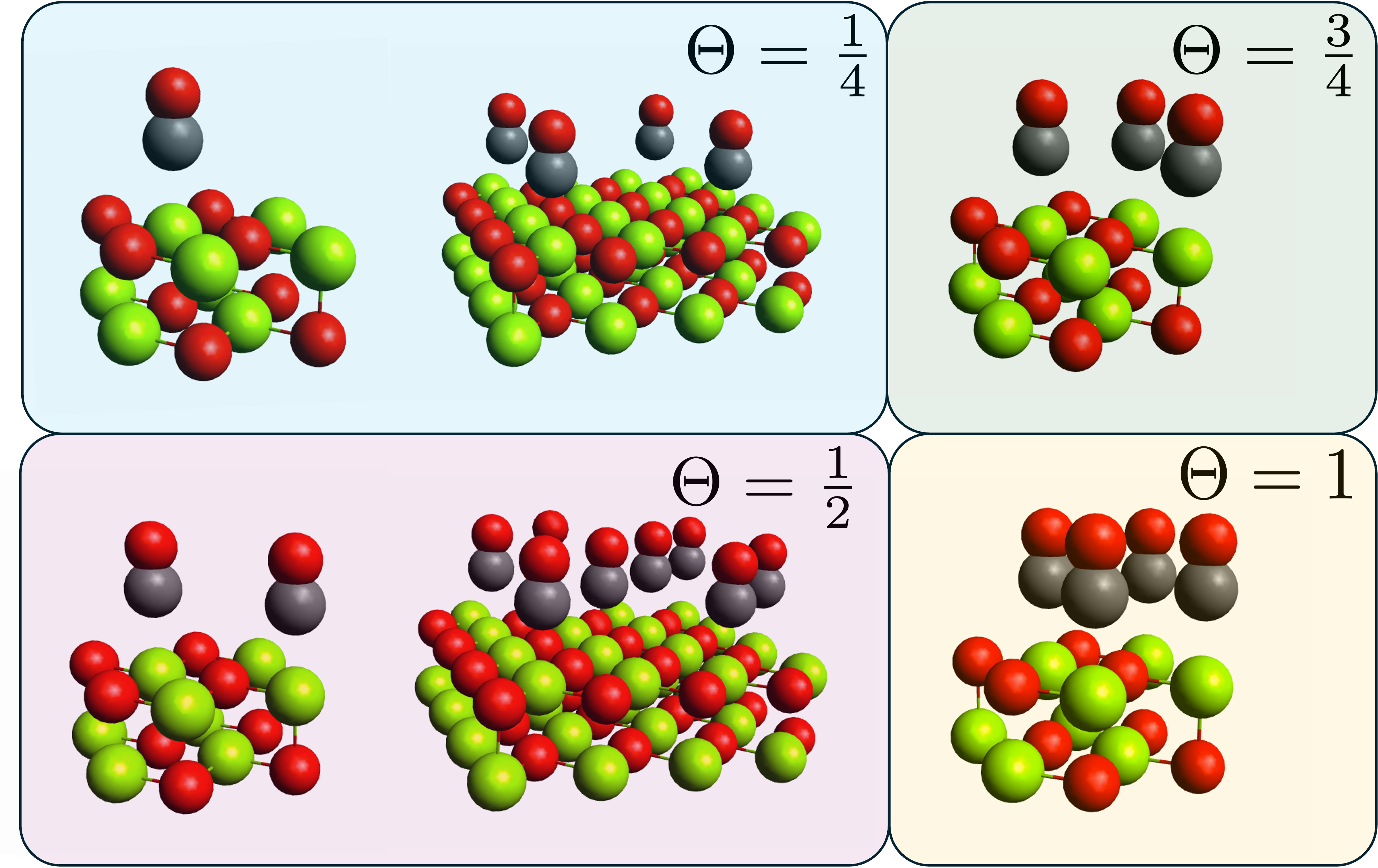}
    \caption{\label{fig:dense}$2\times2$ and $4\times4$ surface slab unit cells used to probe adsorption coverage ratios towards the dense regime.}
\end{figure}

We use a pristine version of the earlier $2\cdot2$ surface slab unit cell, employing the same Mg-C interaction distance as before. Since no geometry optimization has been performed, $\Delta_{\mathrm{geom}}$ is zero. We populate the unit cells with one, two (separated diagonally), three and four CO adsorption sites, in order to capture coverage ratios of $\frac{1}{4}$, $\frac{1}{2}$, $\frac{3}{4}$ and 1, respectively. We also utilize a pristine $4\cdot4$ surface slab unit cell, with 4 and 8 COs adsorption sites, to again probe $\frac{1}{4}$ and $\frac{1}{2}$ surface coverages, to verify the consistency of the $E_{\mathrm{ads}}$ calculations using Megacell-DLPNO-MP2. All unit cells employed are shown in Figure \ref{fig:dense}. A supercell size of $3\times3$ is used throughout, having confirmed the thermodynamic limit is correctly probed from Section \ref{sec:diluteregimeres}. We use the same modified `DZ' and `TZ' basis sets as before. The calculations featuring 8 COs on a 4$\cdot$4 surface represent the costliest calculations in this contribution, featuring just shy of 30000 basis functions in the $3\times3$ supercell, with calculation wall times of three days.

With multiple CO adsorbates now in a given unit cell, the expression to evaluate the adsorption energy now needs to be modified,
\begin{align}
    &E_{\mathrm{ads}} =  E_{\mathrm{int}} = - E^{\mathrm{mol}}_{[\mathrm{CO,\overline{CO}}]} +   \nonumber\\
    &\frac{E^{\mathrm{crys}}_{[\mathrm{MgOCO}]}-E^{\mathrm{crys}}_{[\mathrm{MgO,\overline{CO}}]}- 
    E^{\mathrm{crys}}_{[\mathrm{CO,\overline{MgO}}]} + E^{\mathrm{crys}}_{[\mathrm{CO}]}}{n_{\mathrm{CO}}} \label{eq:Eintcohgeneral}
\end{align}
where the first equality is due to the lack of geometry relaxation energy in this specific scheme. $n_{\mathrm{CO}}$ is the number of COs per unit cell, and $E^{\mathrm{mol}}_{[\mathrm{CO,\bar{CO}}]}$ is the energy of a single CO molecule, with ghost functions representing all remaining COs spanning the reference unit cell and the neighboring cells.

\subsection{Dense Coverage Results}
First, to confirm the validity of Megacell-DLPNO-MP2, we verify whether different unit cell choices, at a given coverage ratio, reproduce the same adsorption energy value. Again, we emphasize that the $3\times3$ supercell sizes used are crucial in accurately representing the thermodynamic bulk of the surface. This ensures different unit cell surface slabs can give the same interaction energy. In Table \ref{table:densetol2244}, MP2 correlation energy values for $E_{\mathrm{ads}}$ at the same coverage ratios are given, for the $2\cdot2$ and $4\cdot4$ surface unit cells. Significant deviations, on the order of $2\;\mathrm{kJ \; mol^{-1}}$, exist for PNO truncation thresholds of $10^{-6}$, but rapidly reduce to within $0.1\;\mathrm{kJ \; mol^{-1}}$ for $10^{-8}$. From this, we can conclude that employing PNO truncation thresholds up to $10^{-8}$ is necessary to obtain robust agreement with larger unit cells. HF energies are also presented, with no significant differences in energies between unit cells. 

\begin{table}[tbp]
    \centering
    \caption{Hartree--Fock and MP2 correlation energies for the adsorption energy ($E_\mathrm{ads}$, in kJ\,mol$^{-1}$), comparing the $2{\cdot}2$ and $4{\cdot}4$ surface slab unit cells, at $\frac{1}{4}$ and $\frac{1}{2}$ coverage ratios. All calculations used the `TZ' basis set. Different PNO truncation thresholds ($\mathcal{T}_\mathrm{PNO}=10^{-X}$, $X=6,7,8$), including the complete PNO space extrapolation, are given.}
    \label{table:densetol2244}
    \begin{ruledtabular}
    \begin{tabular}{lc c d{5}d{5}d{5}c}
    & &  & \multicolumn{3}{c}{DLPNO-MP2} \\
    \cline{4-7}
    \multicolumn{1}{l}{$\Theta$} & \multicolumn{1}{l}{Unit cell} &HF& \multicolumn{1}{c}{6} & \multicolumn{1}{c}{7} & \multicolumn{1}{c}{8} & \multicolumn{1}{c}{CPS} \\
    \hline
    \multicolumn{1}{l}{$1/4$} & \multicolumn{1}{c}{$2{\cdot}2$}  &5.4& -20.0 & -19.1 & -19.9 & -20.3 \\
    & \multicolumn{1}{c}{$4{\cdot}4$} &5.4& -22.0 & -19.5 & -19.9 & -20.1 \\
    \hline
    \multicolumn{1}{l}{$1/2$} & \multicolumn{1}{c}{$2{\cdot}2$}  &7.3& -22.4 & -21.9 & -22.5 & -22.7 \\
    & \multicolumn{1}{c}{$4{\cdot}4$}  &7.3& -23.3 & -22.0 & -22.4 & -22.6 \\
    \end{tabular}
    \end{ruledtabular}
\end{table}

\begin{figure}[htbp]
    \centering
    \includegraphics[scale=0.50]{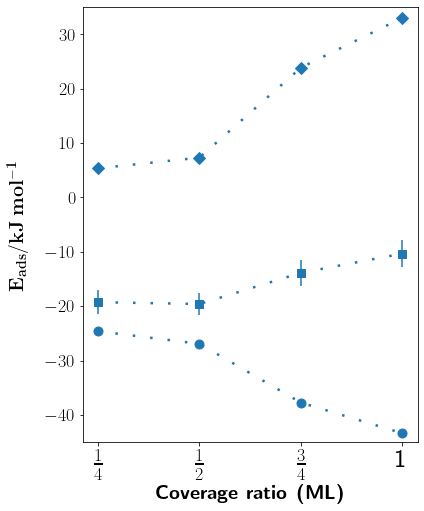}
    \caption{Total CO adsorption energies on the MgO surface at varying coverage ratios up to full monolayer coverage, using the $2\cdot2$ surface slab unit cell, including error bars. Hartree--Fock contributions, in the `TZ' basis, are plotted as diamonds. MP2 correlation energies contributions, extrapolated to the complete PNO space and basis set limits, are plotted as circles.}
    \label{fig:Eint-coh-dense}
\end{figure}

Figure \ref{fig:Eint-coh-dense} presents the adsorption energies of CO on the MgO surface, as a function of coverage ratio, up to full monolayer ($\Theta=1$). The contributions from HF, using the `TZ' basis, and MP2 correlation, at the CPS and basis set limit, are plotted as diamonds and circles, respectively. Pleasingly, the simulated adsorption energy values show an overall decrease in magnitude from $\Theta=\frac{1}{4}$ to $\Theta=1$, with the least exothermic value of $-10.3\;\mathrm{kJ \; mol^{-1}}$ at full monolayer coverage. This reflects the increased lateral repulsion between the CO molecules at higher ratios. One notes the larger increase of the HF repulsion, compared to the attractive dispersive correlation contribution, as the source of this overall behavior. In particular, we can attribute the marked increase in adsorption energy from $\Theta=\frac{1}{2}$ to $\Theta=\frac{3}{4}$ directly to the inclusion of nearest neighbor CO repulsions, which are present in the $\Theta=\frac{3}{4},1$ unit cells (see Figure \ref{fig:dense}). In comparison, we can reason that the diagonal neighbor CO lateral repulsions are much less significant, given the similar adsorption energies between $\Theta=\frac{1}{4}$ to $\Theta=\frac{1}{2}$. Unfortunately, the significant error bars, on the magnitude of around $2\;\mathrm{kJ \; mol^{-1}}$, reiterates our current basis set limitations, and highlights that further work is needed for all-electron basis sets for periodic systems. 

Again, we reiterate caution in directly comparing to the TPD-derived adsorption energy curves presented by Dohn\'alek et al.\cite{dohnalek_physisorption_2001}. As mentioned before, our calculations do not consider non-perpendicular orientations, or clusters of adsorbate molecules. Despite this, we do report a similar adsorption energy at full monolayer coverage, of around $-10\;\mathrm{kJ \; mol^{-1}}$, compared to Figure 4 in Ref.\inlinecite{dohnalek_physisorption_2001}, although admittedly we have not simulated the c(4x2) phase that the authors report to be present. Further to this, the DLPNO-MP2 energy at $\Theta=\frac{3}{4}$ is notably smaller in magnitude than the experimental plot. We can posit that no nearest neighbor adsorption configurations can be responsible for the experimentally-derived adsorption energies between $\Theta=\frac{1}{2}$ to $\frac{3}{4}$, since our calculations report significantly smaller magnitude adsorption energies. Overall, the qualitative behavior of the lateral repulsions, captured entirely within a consistent level of correlated wavefunction theory, indicates promise in simulating unit cells featuring multiple adsorption sites. With further improvements to Megacell-DLPNO-MP2, realistic simulation of even more sophisticated geometries appears feasible.

To quantify the impact of CO-CO interactions, we compute the adsorption energy without lateral contributions, $E'_{\mathrm{int}}$, and compare it directly to the full interactions energies, $E_{\mathrm{int}}$. The $E'_{\mathrm{int}}$ values are shown to be insensitive across coverage ratios, showing that inclusion of lateral interactions is crucial in order to correctly describe adsorption at denser regimes. From the insensitivity of $E'_{\mathrm{int}}$, we can also infer that that the difference in lateral repulsion between the CO monomers in the gas phase, compared to with the influence of the MgO surface, is negligible. This supports the consensus description of CO on MgO adsorption as a form of physisorption. In turn, it also validates the approach taken in Ref \inlinecite{shi_accurate_2025}, where the gas-phase CO interactions are used to approximate the cohesive contribution to the adsorption energy. However, for stronger adsorption interactions described by chemisorption\cite{molina_theoretical_2004,bechthold_dft_2015}, we would expect a significant difference in lateral repulsions from the influence of the surface, and in those cases our evaluation of $E'_{\mathrm{int}}$ would reveal insight into such character.

\begin{table}[htbp]
    \centering
    \caption{Comparison of the adsorption energy (in kJ\,mol$^{-1}$)  at different coverage ratios $\Theta$, with and without gas phase CO lateral interactions. All values are extrapolated to the complete PNO space and basis set limits.}
    \label{tab:E_int}
    \begin{ruledtabular}
    \begin{tabular}{ccc}
    $\Theta$ & $E'_\mathrm{int}$ &  $E_\mathrm{int}$ \\
    \hline
    $1/4$  & -18.9$\pm$ 2.2 & -19.3$\pm$ 2.2\\ 
    $1/2$  & -17.5$\pm$ 2.0 & -19.6$\pm$ 2.1 \\
    $3/4$  & -18.4$\pm$ 1.8 & -13.9$\pm$ 2.4\\
    $1$  & -17.7$\pm$ 1.6 & -10.3$\pm$ 2.5\\
    \end{tabular}
    \end{ruledtabular}
\end{table}

\section{Conclusions}
Correlated wavefunction approaches have been increasingly applied to study adsorption problems, of which CO adsorption on the MgO(001) surface is one of the most fundamental and commonly used case studies. Prior to this work, cluster and periodic wavefunction methods employing MP2 and coupled cluster theory have established excellent agreement with experimentally-derived values for the adsorption energy of a single CO molecule onto a pristine MgO surface. In this work, we leverage the computational efficiency of periodic DLPNO-MP2 to push the scale of correlated wavefunction simulations on adsorption systems, modeling multiple unit cells featuring CO adsorption on MgO. In doing so, we demonstrate the surface slabs in our supercells are converged towards the thermodynamic limit, removing finite-size effects that enable accurate evaluation of adsorption energies across different coverage ratios.

We successfully demonstrate similar estimates of the dilute regime adsorption energy ($\sim -20 \; \mathrm{kJ \; mol^{-1}}$), at three different coverage ratios, compared to previous computational and experimental approaches, validating the accuracy of Megacell-DLPNO-MP2. Tackling the adsorption energy of denser coverages is an inherently trickier problem, due to the vastly larger configuration space that needs to be considered. Here, we employ a simple model consisting of perpendicularly adsorbed CO molecules, evaluating adsorption energies as a function of monolayer coverage ratio, considering the effect of the lateral repulsion between COs. We achieve agreement with experiment of decreased exothermic behavior at higher coverage densities, approaching full monolayer coverage. Our current limitations in basis set contribute the major source of error, and highlights the continued need for improved gaussian basis sets for periodic systems\cite{peintinger_consistent_2013,vilela_oliveira_bssecorrection_2019,klahn_completeness_1977,ye_correlation-consistent_2022}. Overall, however, we show that periodic DLPNO-MP2 is a scalable and efficient approach to model a range of adsorption coverages at the thermodynamic limit, within an entirely consistent level of correlated wavefunction theory.

Whilst the agreement between quantum chemistry and experimental methods for CO on MgO adsorption is impressive, further work is required to model realistic surface systems pertinent to heterogeneous catalysis. Surfaces are not completely pristine in true chemical conditions, and the presence of terraces, defects and kinks on the surface are now understood to have significant influences on the binding with adsorbates\cite{guo_dft_2011, nolan_molecular_2009, lousada_oxygen_2015, camarillo-cisneros_steps_2015, starr_large_2008, arnadottir_adsorption_2010, wandelt_properties_1991, ovcharenko_water_2016}. Dohn{\'a}lek et al.\cite{dohnalek_physisorption_2001}, for example, report significantly more exothermic adsorption energies associated with defect-related sites, which become the dominant mechanism at extremely low coverages. Correlated wavefunction methods need to continue to improve to model increasingly large unit cells if these effects are to be captured. Whilst approaches using DFT are common\cite{nasluzov_cluster_2001,fampiou_binding_2012,lim_dft-based_2011,mehmood_comparative_2009,petersen_revisiting_2017,liu_general_2003,branda_no_2004,kolb_density_2014}, to our knowledge there are significantly fewer studies employing correlated wavefunction methods. In this regard, we believe future versions of periodic DLPNO-MP2 will be a viable solution to explore sophisticated adsorption systems, treating periodic monolayer, defect, kink or terraced surfaces all on equal footing, thanks to its computational efficiency. Work is currently ongoing to model different surface adsorbate systems, moving beyond the weakly physisorbed interactions dominant in CO on MgO, and ongoing method development towards a periodic PNO-CCSD(T) implementation will provide even greater accuracies for simulating systems of relevance to heterogeneous catalysis.

\section*{Acknowledgments}
Financial support for AZ from the University of Oxford and Turbomole GmbH is gratefully acknowledged. PK gratefully acknowledges funding through the Institute for the Promotion of Teaching Science and Technology. AN gratefully acknowledges funding through a Walter Benjamin Fellowship by the Deutsche Forschungsgemeinschaft (DFG, German Research Foundation) -- 517466522.
\bibliography{mgoco_paper.bib}

\end{document}


\fancypagestyle{titlepage}{\fancyhf{}\renewcommand{\headrulewidth}{0pt}\cfoot{\thepage}}
\pagestyle{titlepage}
\begin{center}
\Large
\textbf{Supplementary Material: \\ \ \\ Surface adsorption at the thermodynamic limit using periodic DLPNO-MP2}
\\ \ \\
\normalsize
Andrew Zhu,
Poramas Komonvasee,
Arman Nejad,
and
David P. Tew$^*$
\\ \ \\
\footnotesize \textit{University of Oxford, South Parks Road, Oxford, OX1 3QZ, UK.}
\footnotesize \textit{E-mail: \url{david.tew@chem.ox.ac.uk}}
\\ \ \\
\end{center}
%
%
\section{Unit cell geometries}

The geometries and cell parameters are reported in the Turbomole format
\begin{lstlisting}[language=bash]
$cell
  a b c (* $\alpha$ *) (* $\beta$ *) (* $\gamma$ *)
\end{lstlisting}
where all lengths are in Bohr and angles in degree.

\subsection*{$\textbf{Dilute Coverage Regime}$}
The geometries in this section are taken from the work by Ye and Berkelbach\cite{ye_adsorption_2024}.

\subsubsection*{$\mathrm{1\,CO\, on \, 2\times2 \, MgO \,surface \, (\Theta =0.25)}$}

\begin{lstlisting}[language=bash]
$cell
  11.285742096   11.285742096  90
$coord
        2.8214488818          2.8214134306         10.6032541411      c
        2.8215257748          2.8214088385         12.7588060365      o        
        0.0000000000          0.0000000000          1.8897261246      mg
        0.0000000000          5.6428714587          1.8897261246      mg
        5.6428714587          0.0000000000          1.8897261246      mg
        5.6428714587          5.6428714587          1.8897261246      mg
        2.8214008261          2.8214335562          5.9545282092      mg
        2.8214754136          8.4642984197          5.8611450306      mg
        8.4642871570          2.8214467842          5.8611510022      mg
        8.4643301293          8.4642878751          5.8640625032      mg
        2.8214348034          2.8214348034          1.8897261246      o
        2.8214348034          8.4643062621          1.8897261246      o
        8.4643062621          2.8214348034          1.8897261246      o
        8.4643062621          8.4643062621          1.8897261246      o
       -0.0128432968         -0.0128456968          5.9643583000      o
       -0.0127778178          5.6556489364          5.9651761167      o
        5.6556236329         -0.0127849610          5.9651417426      o
        5.6557114674          5.6556888285          5.9643673518      o
\end{lstlisting}

\subsubsection*{$\mathrm{1\,CO\, on \, 3\times3 \, MgO \,surface \, (\Theta =\frac{1}{9})}$}

\begin{lstlisting}[language=bash]
$cell
16.928613144   16.928613144  90
$coord
    2.82143947100866      2.82143603170712     10.5932477200399       c
    2.82143408528921      2.82143569155641     12.7487996155071       o    
    0.00000000000000      0.00000000000000      1.88972612462577      mg
   11.28574291740374     11.28574291740374      1.88972612462577      mg
   11.28574291740374      5.64287145870187      1.88972612462577      mg
   11.28574291740374      0.00000000000000      1.88972612462577      mg
    5.64287145870187     11.28574291740374      1.88972612462577      mg
    5.64287145870187      5.64287145870187      1.88972612462577      mg
    5.64287145870187      0.00000000000000      1.88972612462577      mg
    0.00000000000000      5.64287145870187      1.88972612462577      mg
    0.00000000000000     11.28574291740374      1.88972612462577      mg
    2.82142628072031     14.10633049987543      5.87026265681236      mg
    2.82145848165348      2.82142972002186      5.94452178813993      mg
   14.10580923782121      8.46566626018438      5.87454679821772      mg
   14.10630833338799      2.82144534805691      5.87031127946555      mg
    2.82141044481539      8.46515006259618      5.87026203320274      mg
    8.46566004298543      8.46566801762967      5.87455159812207      mg
   14.10580931341026     14.10581426449270      5.87458729504857      mg
    8.46519772148904      2.82142473114489      5.87031214873957      mg
    8.46567294981486     14.10579508377254      5.87456881352707      mg
   14.10717772078888     14.10717772078888      1.88972612462577       o
   14.10717772078888      8.46430626208701      1.88972612462577       o
   14.10717772078888      2.82143480338514      1.88972612462577       o
    8.46430626208701     14.10717772078888      1.88972612462577       o
    8.46430626208701      2.82143480338514      1.88972612462577       o
    8.46430626208701      8.46430626208701      1.88972612462577       o
    2.82143480338514      2.82143480338514      1.88972612462577       o
    2.82143480338514     14.10717772078888      1.88972612462577       o
    2.82143480338514      8.46430626208701      1.88972612462577       o
    5.64146945199275     11.28573879780079      5.97146467174535       o
    0.00138955341396     11.28574236938317      5.97146881024556       o
   -0.01728900982789      5.66014661683727      5.95918965334666       o
    5.66014217598088      5.66013616665180      5.95921051592307       o
   11.28574781179441      0.00139898314732      5.97149403808932       o
   -0.01728005252606     -0.01727003697760      5.95921187652588       o
   11.28573961038303      5.64147226768468      5.97148557211628       o
    5.66016054411881     -0.01728058164938      5.95919178873718       o
   11.28574278512292     11.28574040406800      5.96921907239693       o
\end{lstlisting}

\subsubsection*{$\mathrm{1\,CO\, on \, 4\times4 \, MgO \,surface \, (\Theta =\frac{1}{16})}$}

\begin{lstlisting}[language=bash]
$cell
22.571484192 22.571484192  90
$coord
    2.82144204103619      2.82144283472116     10.6034772988813        c
    2.82145479668753      2.82143809150859     12.7588845925054        o    
    0.00000000000000      0.00000000000000      1.88972612462577      mg
   16.92861435720836     11.28574291740374      1.88972612462577      mg
    5.64287145870187     11.28574291740374      1.88972612462577      mg
    5.64287145870187     16.92861435720836      1.88972612462577      mg
   11.28574291740374      0.00000000000000      1.88972612462577      mg
   11.28574291740374      5.64287145870187      1.88972612462577      mg
   11.28574291740374     11.28574291740374      1.88972612462577      mg
   16.92861435720836      5.64287145870187      1.88972612462577      mg
   11.28574291740374     16.92861435720836      1.88972612462577      mg
   16.92861435720836      0.00000000000000      1.88972612462577      mg
    5.64287145870187      5.64287145870187      1.88972612462577      mg
   16.92861435720836     16.92861435720836      1.88972612462577      mg
    0.00000000000000     16.92861435720836      1.88972612462577      mg
    0.00000000000000      5.64287145870187      1.88972612462577      mg
    0.00000000000000     11.28574291740374      1.88972612462577      mg
    5.64287145870187      0.00000000000000      1.88972612462577      mg
    2.82142597836413      2.82143181761786      5.95475136698132      mg
   14.10717928926156     19.75085136823066      5.87976684538355      mg
   14.10717775858340     14.10717758850805      5.87944808638085      mg
    2.82142306818590      8.46820909455494      5.86880201190159      mg
   14.10717711607652      8.46350173008671      5.87976625956846      mg
    8.46816965597072      2.82143074047397      5.86879371600391      mg
   19.74704077329487      8.46730510626870      5.87627321421065      mg
    2.82143661752221     14.10717656805595      5.88008698388633      mg
   19.74705704383680     19.74706511296736      5.87626338763480      mg
    8.46731264627594     19.74705305651468      5.87627081425847      mg
    8.46350781500483     14.10717936485061      5.87976699656164      mg
    2.82144136073479     19.74614211403631      5.86880598032645      mg
   19.75084685178522     14.10717792865876      5.87977071932211      mg
    8.46729834104917      8.46730357559054      5.87626478603214      mg
   14.10718083883699      2.82143476559061      5.88009159481807      mg
   19.74617340790093      2.82144462996098      5.86878638386654      mg
   19.75004916059349      2.82143480338514      1.88972612462577       o
    8.46430626208701     19.75004916059349      1.88972612462577       o
   14.10717772078888      2.82143480338514      1.88972612462577       o
    8.46430626208701      8.46430626208701      1.88972612462577       o
   19.75004916059349      8.46430626208701      1.88972612462577       o
   14.10717772078888     14.10717772078888      1.88972612462577       o
   14.10717772078888     19.75004916059349      1.88972612462577       o
   14.10717772078888      8.46430626208701      1.88972612462577       o
    8.46430626208701     14.10717772078888      1.88972612462577       o
    2.82143480338514     19.75004916059349      1.88972612462577       o
   19.75004916059349     14.10717772078888      1.88972612462577       o
    2.82143480338514      8.46430626208701      1.88972612462577       o
    2.82143480338514     14.10717772078888      1.88972612462577       o
    2.82143480338514      2.82143480338514      1.88972612462577       o
    8.46430626208701      2.82143480338514      1.88972612462577       o
   19.75004916059349     19.75004916059349      1.88972612462577       o
   16.92821176995477     11.28614093152012      5.97052243540235       o
   16.92645859543991      5.64229084035008      5.97167673681105       o
   16.92646305519356      0.00058313168754      5.97167409119448       o
    5.64229652842572     16.92645549628906      5.97167395891365       o
   11.28614183858866     11.28614285904076      5.97051612371709       o
   11.28789359580892      5.64229084035008      5.97166948026273       o
   11.28789660047346      0.00057880421471      5.97166929129012       o
    5.64228980100071     11.28790072007641      5.97167448803696       o
    5.65983263884167      5.65984760547257      5.96179482978207       o
    5.65984297564357     -0.01698243955870      5.96180388157021       o
    0.00057831288592     16.92645825528920      5.97167547069455       o
    0.00057555388578     11.28790036102845      5.97167796513303       o
   -0.01698468833279      5.65985584467848      5.96178908501465       o
   -0.01697070435946     -0.01696960831831      5.96179919504942       o
   11.28614478656141     16.92821424549599      5.97051901499806       o
   16.92821396203707     16.92821634309199      5.97051721975824       o

\end{lstlisting}
\subsection*{$\textbf{Dense Coverage Regime}$}
The geometries in this section are CO molecules adsorbed on a pristine MgO surface without geometry optimization. 

\subsubsection*{$\mathrm{1\,CO\, on \, 2\times2 \, MgO \,surface \, (\Theta =0.25)}$}

\begin{lstlisting}[language=bash]
$cell
11.285742096   11.285742096  90
$coord
 -8.4643072924     14.1071768992      7.12154955     c
 -8.4643072924     14.1071768992      9.277101446    o
-14.1071783403     14.1071768992      2.4728236184   mg   
-14.1071783403     19.7500479471      2.4728236184   mg 
-11.2857420958     11.2857420958     -1.5186979106   mg    
 -8.4643072924     14.1071768992      2.4728236184   mg
-11.2857420958     16.9286131437     -1.5186979106   mg    
 -8.4643072924     19.7500479471      2.4728236184   mg  
 -5.6428710479     11.2857420958     -1.5186979106   mg  
 -5.6428710479     16.9286131437     -1.5186979106   mg 
-14.1071783403     14.1071768992     -1.5186979106    o  
-14.1071783403     19.7500479471     -1.5186979106    o 
 -8.4643072924     14.1071768992     -1.5186979106    o    
-11.2857420958     11.2857420958      2.5645722028    o    
 -8.4643072924     19.7500479471     -1.5186979106    o    
-11.2857420958     16.9286131437      2.5645722028    o    
 -5.6428710479     11.2857420958      2.5645722028    o    
 -5.6428710479     16.9286131437      2.5645722028    o  
\end{lstlisting}

\subsubsection*{$\mathrm{2\,CO\, on \, 2\times2 \, MgO \,surface \, (\Theta =0.50)}$}

\begin{lstlisting}[language=bash]
$cell
11.285742096   11.285742096  90
$coord
-14.1071783403     19.7500479471      7.12154955     c
 -8.4643072924     14.1071768992      7.12154955     c
-14.1071783403     19.7500479471      9.277101446    o
 -8.4643072924     14.1071768992      9.277101446    o
-14.1071783403     14.1071768992      2.4728236184   mg   
-14.1071783403     19.7500479471      2.4728236184   mg 
-11.2857420958     11.2857420958     -1.5186979106   mg    
 -8.4643072924     14.1071768992      2.4728236184   mg
-11.2857420958     16.9286131437     -1.5186979106   mg    
 -8.4643072924     19.7500479471      2.4728236184   mg  
 -5.6428710479     11.2857420958     -1.5186979106   mg  
 -5.6428710479     16.9286131437     -1.5186979106   mg 
-14.1071783403     14.1071768992     -1.5186979106    o  
-14.1071783403     19.7500479471     -1.5186979106    o 
 -8.4643072924     14.1071768992     -1.5186979106    o    
-11.2857420958     11.2857420958      2.5645722028    o    
 -8.4643072924     19.7500479471     -1.5186979106    o    
-11.2857420958     16.9286131437      2.5645722028    o    
 -5.6428710479     11.2857420958      2.5645722028    o    
 -5.6428710479     16.9286131437      2.5645722028    o  
\end{lstlisting}

\subsubsection*{$\mathrm{3\,CO\, on \, 2\times2 \, MgO \,surface \, (\Theta =0.75)}$}

\begin{lstlisting}[language=bash]
$cell
11.285742096   11.285742096  90
$coord
-14.1071783403     19.7500479471      7.12154955      c
 -8.4643072924     14.1071768992      7.12154955      c
-14.1071783403     14.1071768992      7.12154955      c
-14.1071783403     19.7500479471      9.277101446     o
 -8.4643072924     14.1071768992      9.277101446     o
-14.1071783403     14.1071768992      9.277101446     o
-14.1071783403     14.1071768992      2.4728236184   mg   
-14.1071783403     19.7500479471      2.4728236184   mg 
-11.2857420958     11.2857420958     -1.5186979106   mg    
 -8.4643072924     14.1071768992      2.4728236184   mg
-11.2857420958     16.9286131437     -1.5186979106   mg    
 -8.4643072924     19.7500479471      2.4728236184   mg  
 -5.6428710479     11.2857420958     -1.5186979106   mg  
 -5.6428710479     16.9286131437     -1.5186979106   mg 
-14.1071783403     14.1071768992     -1.5186979106    o  
-14.1071783403     19.7500479471     -1.5186979106    o 
 -8.4643072924     14.1071768992     -1.5186979106    o    
-11.2857420958     11.2857420958      2.5645722028    o    
 -8.4643072924     19.7500479471     -1.5186979106    o    
-11.2857420958     16.9286131437      2.5645722028    o    
 -5.6428710479     11.2857420958      2.5645722028    o    
 -5.6428710479     16.9286131437      2.5645722028    o 
\end{lstlisting}

\subsubsection*{$\mathrm{4\,CO\, on \, 2\times2 \, MgO \,surface \, (\Theta =1.00)}$}

\begin{lstlisting}[language=bash]
$cell
11.285742096   11.285742096  90
$coord
-14.1071783403     19.7500479471      7.1215495 5     c
 -8.4643072924     14.1071768992      7.1215495       c
-14.1071783403     14.1071768992      7.12154955      c
-8.4643072924      19.7500479471      7.12154955      c
-14.1071783403     19.7500479471      9.277101446     o
 -8.4643072924     14.1071768992      9.277101446     o
-14.1071783403     14.1071768992      9.277101446     o
-8.4643072924      19.7500479471      9.277101446     o
-14.1071783403     14.1071768992      2.4728236184   mg   
-14.1071783403     19.7500479471      2.4728236184   mg 
-11.2857420958     11.2857420958     -1.5186979106   mg    
 -8.4643072924     14.1071768992      2.4728236184   mg
-11.2857420958     16.9286131437     -1.5186979106   mg    
 -8.4643072924     19.7500479471      2.4728236184   mg  
 -5.6428710479     11.2857420958     -1.5186979106   mg  
 -5.6428710479     16.9286131437     -1.5186979106   mg 
-14.1071783403     14.1071768992     -1.5186979106    o  
-14.1071783403     19.7500479471     -1.5186979106    o 
 -8.4643072924     14.1071768992     -1.5186979106    o    
-11.2857420958     11.2857420958      2.5645722028    o    
 -8.4643072924     19.7500479471     -1.5186979106    o    
-11.2857420958     16.9286131437      2.5645722028    o    
 -5.6428710479     11.2857420958      2.5645722028    o    
 -5.6428710479     16.9286131437      2.5645722028    o  
\end{lstlisting}

\subsubsection*{$\mathrm{4\,CO\, on \, 4\times4 \, MgO \,surface \, (\Theta =0.25)}$}

\begin{lstlisting}[language=bash]
$cell
22.571484192 22.571484192  90
$coord
14.1071783403     -14.1071783403      7.1215495500    c   
 -2.8214362445     -2.8214362445      7.1215495500    c    
 -2.8214362445    -14.1071783403      7.1215495500    c    
-14.1071783403     -2.8214362445      7.1215495500    c    
-14.1071783403    -14.1071783403      9.2771014460    o    
 -2.8214362445     -2.8214362445      9.2771014460    o    
 -2.8214362445    -14.1071783403      9.2771014460    o    
-14.1071783403     -2.8214362445      9.2771014460    o    
-16.9286131437    -16.9286131437     -1.5186979106   mg    
-14.1071783403    -14.1071783403      2.4728236184   mg    
-16.9286131437    -11.2857420958     -1.5186979106   mg    
-14.1071783403     -8.4643072924      2.4728236184   mg    
-16.9286131437     -5.6428710479     -1.5186979106   mg    
-14.1071783403     -2.8214362445      2.4728236184   mg    
-16.9286131437      0.0000000000     -1.5186979106   mg    
-14.1071783403      2.8214348034      2.4728236184   mg    
-11.2857420958    -16.9286131437     -1.5186979106   mg    
 -8.4643072924    -14.1071783403      2.4728236184   mg    
-11.2857420958    -11.2857420958     -1.5186979106   mg    
 -8.4643072924     -8.4643072924      2.4728236184   mg    
-11.2857420958     -5.6428710479     -1.5186979106   mg    
 -8.4643072924     -2.8214362445      2.4728236184   mg    
-11.2857420958      0.0000000000     -1.5186979106   mg    
 -8.4643072924      2.8214348034      2.4728236184   mg    
 -5.6428710479    -16.9286131437     -1.5186979106   mg    
 -2.8214362445    -14.1071783403      2.4728236184   mg    
 -5.6428710479    -11.2857420958     -1.5186979106   mg    
 -2.8214362445     -8.4643072924      2.4728236184   mg    
 -5.6428710479     -5.6428710479     -1.5186979106   mg    
 -2.8214362445     -2.8214362445      2.4728236184   mg    
 -5.6428710479      0.0000000000     -1.5186979106   mg    
 -2.8214362445      2.8214348034      2.4728236184   mg    
  0.0000000000    -16.9286131437     -1.5186979106   mg    
  2.8214348034    -14.1071783403      2.4728236184   mg    
  0.0000000000    -11.2857420958     -1.5186979106   mg    
  2.8214348034     -8.4643072924      2.4728236184   mg    
  0.0000000000     -5.6428710479     -1.5186979106   mg    
  2.8214348034     -2.8214362445      2.4728236184   mg    
  0.0000000000      0.0000000000     -1.5186979106   mg    
  2.8214348034      2.8214348034      2.4728236184   mg    
-14.1071783403    -14.1071783403     -1.5186979106    o    
-16.9286131437    -16.9286131437      2.5645722028    o    
-14.1071783403     -8.4643072924     -1.5186979106    o    
-16.9286131437    -11.2857420958      2.5645722028    o    
-14.1071783403     -2.8214362445     -1.5186979106    o    
-16.9286131437     -5.6428710479      2.5645722028    o    
-14.1071783403      2.8214348034     -1.5186979106    o    
-16.9286131437      0.0000000000      2.5645722028    o    
 -8.4643072924    -14.1071783403     -1.5186979106    o    
-11.2857420958    -16.9286131437      2.5645722028    o    
 -8.4643072924     -8.4643072924     -1.5186979106    o    
-11.2857420958    -11.2857420958      2.5645722028    o    
 -8.4643072924     -2.8214362445     -1.5186979106    o    
-11.2857420958     -5.6428710479      2.5645722028    o    
 -8.4643072924      2.8214348034     -1.5186979106    o    
-11.2857420958      0.0000000000      2.5645722028    o    
 -2.8214362445    -14.1071783403     -1.5186979106    o    
 -5.6428710479    -16.9286131437      2.5645722028    o    
 -2.8214362445     -8.4643072924     -1.5186979106    o    
 -5.6428710479    -11.2857420958      2.5645722028    o    
 -2.8214362445     -2.8214362445     -1.5186979106    o    
 -5.6428710479     -5.6428710479      2.5645722028    o    
 -2.8214362445      2.8214348034     -1.5186979106    o    
 -5.6428710479      0.0000000000      2.5645722028    o    
  2.8214348034    -14.1071783403     -1.5186979106    o    
  0.0000000000    -16.9286131437      2.5645722028    o    
  2.8214348034     -8.4643072924     -1.5186979106    o    
  0.0000000000    -11.2857420958      2.5645722028    o    
  2.8214348034     -2.8214362445     -1.5186979106    o    
  0.0000000000     -5.6428710479      2.5645722028    o    
  2.8214348034      2.8214348034     -1.5186979106    o    
  0.0000000000      0.0000000000      2.5645722028    o
\end{lstlisting}

\subsubsection*{$\mathrm{8\,CO\, on \, 4\times4 \, MgO \,surface \, (\Theta =0.50)}$}

\begin{lstlisting}[language=bash]
$cell
22.571484192 22.571484192  90
$coord
-14.1071783403    -14.1071783403      7.12154955      c
-8.4643072924      -8.4643072924      7.12154955      c
-2.8214362445      -2.8214362445      7.12154955      c
 2.8214348034       2.8214348034      7.12154955      c
-2.8214362445     -14.1071783403      7.12154955      c
2.8214348034       -8.4643072924      7.12154955      c
-14.1071783403     -2.8214362445      7.12154955      c
-8.4643072924       2.8214348034      7.12154955      c
-14.1071783403    -14.1071783403      9.277101446     o
-8.4643072924      -8.4643072924      9.277101446     o
-2.8214362445      -2.8214362445      9.277101446     o
 2.8214348034       2.8214348034      9.277101446     o
-2.8214362445     -14.1071783403      9.277101446     o
2.8214348034       -8.4643072924      9.277101446     o
-14.1071783403     -2.8214362445      9.277101446     o
-8.4643072924       2.8214348034      9.277101446     o
-16.9286131437    -16.9286131437     -1.5186979106   mg    
-14.1071783403    -14.1071783403      2.4728236184   mg
-16.9286131437    -11.2857420958     -1.5186979106   mg    
-14.1071783403     -8.4643072924      2.4728236184   mg  
-16.9286131437     -5.6428710479     -1.5186979106   mg    
-14.1071783403     -2.8214362445      2.4728236184   mg 
-16.9286131437      0.0000000000     -1.5186979106   mg    
-14.1071783403      2.8214348034      2.4728236184   mg 
-11.2857420958    -16.9286131437     -1.5186979106   mg    
 -8.4643072924    -14.1071783403      2.4728236184   mg 
-11.2857420958    -11.2857420958     -1.5186979106   mg    
 -8.4643072924     -8.4643072924      2.4728236184   mg 
-11.2857420958     -5.6428710479     -1.5186979106   mg    
 -8.4643072924     -2.8214362445      2.4728236184   mg 
-11.2857420958      0.0000000000     -1.5186979106   mg    
 -8.4643072924      2.8214348034      2.4728236184   mg   
 -5.6428710479    -16.9286131437     -1.5186979106   mg    
 -2.8214362445    -14.1071783403      2.4728236184   mg 
 -5.6428710479    -11.2857420958     -1.5186979106   mg    
 -2.8214362445     -8.4643072924      2.4728236184   mg  
 -5.6428710479     -5.6428710479     -1.5186979106   mg    
 -2.8214362445     -2.8214362445      2.4728236184   mg  
 -5.6428710479      0.0000000000     -1.5186979106   mg    
 -2.8214362445      2.8214348034      2.4728236184   mg  
  0.0000000000    -16.9286131437     -1.5186979106   mg    
  2.8214348034    -14.1071783403      2.4728236184   mg 
  0.0000000000    -11.2857420958     -1.5186979106   mg    
  2.8214348034     -8.4643072924      2.4728236184   mg 
  0.0000000000     -5.6428710479     -1.5186979106   mg    
  2.8214348034     -2.8214362445      2.4728236184   mg  
  0.0000000000      0.0000000000     -1.5186979106   mg    
  2.8214348034      2.8214348034      2.4728236184   mg  
-14.1071783403    -14.1071783403     -1.5186979106    o    
-16.9286131437    -16.9286131437      2.5645722028    o    
-14.1071783403     -8.4643072924     -1.5186979106    o    
-16.9286131437    -11.2857420958      2.5645722028    o    
-14.1071783403     -2.8214362445     -1.5186979106    o    
-16.9286131437     -5.6428710479      2.5645722028    o    
-14.1071783403      2.8214348034     -1.5186979106    o    
-16.9286131437      0.0000000000      2.5645722028    o    
 -8.4643072924    -14.1071783403     -1.5186979106    o    
-11.2857420958    -16.9286131437      2.5645722028    o    
 -8.4643072924     -8.4643072924     -1.5186979106    o    
-11.2857420958    -11.2857420958      2.5645722028    o    
 -8.4643072924     -2.8214362445     -1.5186979106    o    
-11.2857420958     -5.6428710479      2.5645722028    o    
 -8.4643072924      2.8214348034     -1.5186979106    o    
-11.2857420958      0.0000000000      2.5645722028    o    
 -2.8214362445    -14.1071783403     -1.5186979106    o    
 -5.6428710479    -16.9286131437      2.5645722028    o    
 -2.8214362445     -8.4643072924     -1.5186979106    o    
 -5.6428710479    -11.2857420958      2.5645722028    o    
 -2.8214362445     -2.8214362445     -1.5186979106    o    
 -5.6428710479     -5.6428710479      2.5645722028    o    
 -2.8214362445      2.8214348034     -1.5186979106    o    
 -5.6428710479      0.0000000000      2.5645722028    o    
  2.8214348034    -14.1071783403     -1.5186979106    o    
  0.0000000000    -16.9286131437      2.5645722028    o    
  2.8214348034     -8.4643072924     -1.5186979106    o    
  0.0000000000    -11.2857420958      2.5645722028    o    
  2.8214348034     -2.8214362445     -1.5186979106    o    
  0.0000000000     -5.6428710479      2.5645722028    o    
  2.8214348034      2.8214348034     -1.5186979106    o    
  0.0000000000      0.0000000000      2.5645722028    o  
\end{lstlisting}
\newpage
\section{$\textbf{Basis sets}$}
\subsection*{$\textbf{Orbital Basis Sets}$}
   \subsubsection*{\textbf{‘DZ’ Basis Set}}
   \begin{lstlisting}[language=bash]
# c     
*
   8  s
  6665.0000000      0.69200000000E-03
  1000.0000000      0.53290000000E-02
  228.00000000      0.27077000000E-01
  64.710000000      0.10171800000
  21.060000000      0.27474000000
  7.4950000000      0.44856400000
  2.7970000000      0.28507400000
 0.52150000000      0.15204000000E-01
   8  s
  6665.0000000     -0.14600000000E-03
  1000.0000000     -0.11540000000E-02
  228.00000000     -0.57250000000E-02
  64.710000000     -0.23312000000E-01
  21.060000000     -0.63955000000E-01
  7.4950000000     -0.14998100000
  2.7970000000     -0.12726200000
 0.52150000000      0.54452900000
   1  s
 0.15960000000       1.0000000000
   3  p
  9.4390000000      0.38109000000E-01
  2.0020000000      0.20948000000
 0.54560000000      0.50855700000
   1  p
 0.15170000000       1.0000000000
   1  d
 0.55000000000       1.0000000000
*
# o     
*
   8  s
  11720.000000      0.71000000000E-03
  1759.0000000      0.54700000000E-02
  400.80000000      0.27837000000E-01
  113.70000000      0.10480000000
  37.030000000      0.28306200000
  13.270000000      0.44871900000
  5.0250000000      0.27095200000
  1.0130000000      0.15458000000E-01
   8  s
  11720.000000     -0.16000000000E-03
  1759.0000000     -0.12630000000E-02
  400.80000000     -0.62670000000E-02
  113.70000000     -0.25716000000E-01
  37.030000000     -0.70924000000E-01
  13.270000000     -0.16541100000
  5.0250000000     -0.11695500000
  1.0130000000      0.55736800000
   1  s
 0.30230000000       1.0000000000
   3  p
  17.700000000      0.43018000000E-01
  3.8540000000      0.22891300000
  1.0460000000      0.50872800000
   1  p
 0.27530000000       1.0000000000
   1  d
  1.1850000000       1.0000000000
*
# mg    
*
   5  s
  4953.8339196     -0.57778967498E-02
  745.18044154     -0.43124761082E-01
  169.21604972     -0.19268216987
  47.300672019     -0.48641439116
  14.461336973     -0.42550894077
   3  s
  24.768174789      0.87956969984E-01
  2.4940945349     -0.55165058128
  0.87807584530    -0.53443294833
   1  s
  0.34506887000     1.0000000000
   1  s
  0.15005399000     1.0000000000
   1  s
  2.911600E+00      1.000000E+00
   5  p
  98.053010494     -0.14480564601E-01
  22.586932277     -0.95495750787E-01
  6.8391509842     -0.30787672651
  2.2332843818     -0.49936292886
  0.71606599390    -0.31503476213
   1  p
  0.24692323000     1.0000000000
   1  p
  6.031000E+00      1.000000E+00
   1  d
  1.932000D-01      1.0000000
   1  d
  1.370200E+00      1.000000E+00
*

    \end{lstlisting}
   \subsubsection*{\textbf{‘TZ’ Basis Set}}
     \begin{lstlisting}[language=bash]
# c     
*
   8  s
  8236.0000000      0.53100000000E-03
  1235.0000000      0.41080000000E-02
  280.80000000      0.21087000000E-01
  79.270000000      0.81853000000E-01
  25.590000000      0.23481700000
  8.9970000000      0.43440100000
  3.3190000000      0.34612900000
 0.36430000000     -0.89830000000E-02
   8  s
  8236.0000000     -0.11300000000E-03
  1235.0000000     -0.87800000000E-03
  280.80000000     -0.45400000000E-02
  79.270000000     -0.18133000000E-01
  25.590000000     -0.55760000000E-01
  8.9970000000     -0.12689500000
  3.3190000000     -0.17035200000
 0.36430000000      0.59868400000
   1  s
 0.90590000000       1.0000000000
   1  s
 0.12850000000       1.0000000000
   3  p
  18.710000000      0.14031000000E-01
  4.1330000000      0.86866000000E-01
  1.2000000000      0.29021600000
   1  p
 0.38270000000       1.0000000000
   1  p
 0.12090000000       1.0000000000
   1  d
  1.0970000000       1.0000000000
   1  d
 0.31800000000       1.0000000000
   1  f
 0.76100000000       1.0000000000
*
# o     
*
   8  s
  15330.000000      0.50800000000E-03
  2299.0000000      0.39290000000E-02
  522.40000000      0.20243000000E-01
  147.30000000      0.79181000000E-01
  47.550000000      0.23068700000
  16.760000000      0.43311800000
  6.2070000000      0.35026000000
 0.68820000000     -0.81540000000E-02
   8  s
  15330.000000     -0.11500000000E-03
  2299.0000000     -0.89500000000E-03
  522.40000000     -0.46360000000E-02
  147.30000000     -0.18724000000E-01
  47.550000000     -0.58463000000E-01
  16.760000000     -0.13646300000
  6.2070000000     -0.17574000000
 0.68820000000      0.60341800000
   1  s
  1.7520000000       1.0000000000
   1  s
 0.23840000000       1.0000000000
   3  p
  34.460000000      0.15928000000E-01
  7.7490000000      0.99740000000E-01
  2.2800000000      0.31049200000
   1  p
 0.71560000000       1.0000000000
   1  p
 0.21400000000       1.0000000000
   1  d
  2.3140000000       1.0000000000
   1  d
 0.64500000000       1.0000000000
   1  f
  1.4280000000       1.0000000000
*
# mg    
*
   7  s
  31438.3495550      0.00060912311326
  4715.51533540      0.00470661964650
  1073.16292470      0.02413582065700
  303.572387680      0.09362895983400
  98.6262510420      0.26646742093000
  34.9438084170      0.47890929917000
  12.8597851990      0.33698490286000
   3  s
  64.8769130040      0.01918088930700
  19.7255207770      0.09091370439200
  2.89518043390     -0.39563756125000
   2  s
  1.19604547100       1.68276033730000
  0.54329451156       0.52141091954000
   1  s
  0.83471188300       1.00000000000000
   1  s
  0.14506887000       1.00000000000000
   1  s
  2.587700E+01        1.000000E+00
   1  s
  3.040200E+00        1.000000E+00
   5  p
  179.871896120      0.00537995490180
  42.1200693760      0.03931801409800
  13.1205030320      0.15740129476000
  4.62575036090      0.35919094128000
  1.66952110160      0.45533379310000
   1  p
  0.56631001000       1.00000000000000
   1  p
  0.18813966000       1.00000000000000
   1  p
  1.705300E+01        1.000000E+00
   1  p
  3.954100E+00        1.000000E+00
   1  d
  0.126000000         1.00000000
   1  d
  0.294000000         1.00000000
   1  d
  4.550700E+00        1.000000E+00
   1  d
  1.105100E+00        1.000000E+00
   1  f
  0.252000000         1.00000000
   1  f
  1.298500E+00        1.000000E+00
    \end{lstlisting}
\subsection*{$\textbf{Auxiliary Basis Set}$}
\subsubsection*{\textbf{Standard Basis Set}}
    The standard basis sets\cite{weigend_accurate_2006} are used for density fitting for non-ghost atoms, as mentioned in the main text.  
\subsubsection*{\textbf{‘Dummy’ Basis Set}}
    These basis sets are used for density fitting for ghost atoms.
     \begin{lstlisting}[language=bash]
c dummy
# c     
*
   1  s
  100000000000000      1.0000000000
*
mg dummy
# mg     
*
   1  s
  100000000000000      1.0000000000
*
o dummy
# o     
*
   1  s
  100000000000000      1.0000000000
*
     \end{lstlisting}

\newpage
\section{$\textbf{Further results}$}
\subsubsection*{\textbf{Dilute coverage results}}
\begin{table*}[htbp]
\centering
    \caption{Hartree--Fock and MP2 correlation energies for each system involved in evaluating $E_{\mathrm{int}}$, comparing supercell sizes of $3\times3$ and $5\times5$. All energies are given in Hartrees, unless stated. Calculations employed the $2\cdot2$ surface slab unit cell, using the modified `TZ' basis set. This data was used for Table I, Table II and Figure 3 of the main text.}
    \label{tab:tdl}
    \begin{tabular}{|l|cc|cc|cc|cc|}
    System  & \multicolumn{2}{c|}{HF} & \multicolumn{2}{c|}{$E_{\mathrm{corr}}(\mathcal{T}_{\mathrm{PNO}}=10^{-7})$} & \multicolumn{2}{c|}{$E_{\mathrm{corr}} (\mathcal{T}_{\mathrm{PNO}}=10^{-8})$}&  \multicolumn{2}{c|}{CPS} \\
    $k_{\mathrm{super}}$& $3^2$ & $5^2$ & $3^2$ & $5^2$ & $3^2$ & $5^2$ & $3^2$ & $5^2$ \\
    \hline
    $[\mathrm{MgOCO}]$        & -2310.10796  & -2310.10796  & -4.14200 & -4.14224 & -4.14370 & -4.14412 & -4.14449 & -4.14499 \\
    $[\mathrm{MgO,\overline{CO}}]$ & -2197.33009  & -2197.33009  & -3.77718 & -3.77739 & -3.77861 & -3.77898 & -3.77927 & -3.77972 \\
    $[\mathrm{CO,\overline{MgO}}]$ &  -112.77868  &  -112.77868  & -0.35807 & -0.35807 & -0.35814 & -0.35814 & -0.35817 & -0.35817 \\
    $[\mathrm{CO}]$           &  -112.77842  &  -112.77842  & -0.35712 & -0.35712 & -0.35714 & -0.35714 & -0.35715 & -0.35715 \\
    $[\mathrm{CO,\overline{CO}}]$  &  -112.77855  &  -112.77855  & -0.35685 & -0.35685 & -0.35687 & -0.35687 & -0.35688 & -0.35688 \\ 
    \hline 
    $E_{\mathrm{int}}$    &     0.00094  &     0.00094  & -0.00702 & -0.00705 & -0.00723 & -0.00727 & -0.00732 & -0.00738 \\
    $E_{\mathrm{int}} (\mathrm{kJ \; mol^{-1}})$ & 2.47 & 2.47 & -18.42 & -18.51  & -18.97 & -19.11 & -19.23  & -19.39 \\
    \end{tabular}
\end{table*}

\begin{table*}[h]
\centering
    \caption{Hartree--Fock and MP2 correlation energies for each system involved in evaluating $E_{\mathrm{int}}$, employing the $3\cdot3$ surface slab unit cell. All energies are given in Hartrees, unless stated. Calculations used a $3\times3$ supercell size, using the modified `TZ' basis set. This data was used for Table II and Figure 3 of the main text.}
    \label{tab:tdl}
    \begin{tabular}{|l|c|c|c|c|}
    System  & \multicolumn{1}{c|}{HF} & \multicolumn{1}{c|}{$E_{\mathrm{corr}}(\mathcal{T}_{\mathrm{PNO}}=10^{-7})$} & \multicolumn{1}{c|}{$E_{\mathrm{corr}} (\mathcal{T}_{\mathrm{PNO}}=10^{-8})$}&  \multicolumn{1}{c|}{CPS} \\
    \hline
    $[\mathrm{MgOCO}]$          & -5056.76941   & -8.86202 & -8.86541  & -8.86698 \\
    $[\mathrm{MgO,\overline{CO}}]$   & -4943.99143   & -8.49714  & -8.50039  & -8.50189 \\
    $[\mathrm{CO,\overline{MgO}}]$   &  -112.77878   & -0.35784 & -0.35791  & -0.35794 \\
    $[\mathrm{CO}]$             &  -112.77852   & -0.35689  & -0.35691  & -0.35693 \\
    $[\mathrm{CO,\overline{CO}}]$    &  -112.77855   & -0.35685  & -0.35687  & -0.35688 \\ 
    \hline 
    $E_{\mathrm{int}}$      &     0.00083 & -0.00708 & -0.00716  & -0.00719 \\
    $E_{\mathrm{int}} (\mathrm{kJ \; mol^{-1}})$  & 2.17 & -18.59   & -18.79  & -18.89 \\
    \end{tabular}
\end{table*}

\begin{table*}[h]
\centering
    \caption{Hartree--Fock and MP2 correlation energies for each system involved in evaluating $E_{\mathrm{int}}$, employing the $4\cdot4$ surface slab unit cell. All energies are given in Hartrees, unless stated. Calculations used a $3\times3$ supercell size, using the modified `TZ' basis set. This data was used for Table II and Figure 3 of the main text.}
    \label{tab:tdl}
    \begin{tabular}{|l|c|c|c|c|}
    System  & \multicolumn{1}{c|}{HF} & \multicolumn{1}{c|}{$E_{\mathrm{corr}}(\mathcal{T}_{\mathrm{PNO}}=10^{-7})$} & \multicolumn{1}{c|}{$E_{\mathrm{corr}} (\mathcal{T}_{\mathrm{PNO}}=10^{-8})$}&  \multicolumn{1}{c|}{CPS} \\
    \hline
    $[\mathrm{MgOCO}]$          & -8902.09505   & -15.46972 & -15.47562  & -15.47835 \\
    $[\mathrm{MgO,\overline{CO}}]$   & -8789.31697   & -15.10511  & -15.11063  & -15.11318 \\
    $[\mathrm{CO,\overline{MgO}}]$   &  -112.77881   & -0.35778 & -0.35785  & -0.35789 \\
    $[\mathrm{CO}]$             &  -112.77855   & -0.35684  & -0.35686  & -0.35687 \\
    $[\mathrm{CO,\overline{CO}}]$    &  -112.77856   & -0.35684  & -0.35686  & -0.35687 \\ 
    \hline 
    $E_{\mathrm{int}}$      &     0.00074 &-0.00683 & -0.00714  & -0.00729 \\
    $E_{\mathrm{int}} (\mathrm{kJ \; mol^{-1}})$  & 1.94 & -17.92   & -18.75  & -19.13 \\
    \end{tabular}
\end{table*}
\clearpage
\newpage
\subsubsection*{\textbf{Dense coverage results}}

\begin{table*}[h]
\centering
    \caption{Hartree--Fock and MP2 correlation energies for each system involved in evaluating $E_{\mathrm{int}}$, employing the $2\cdot2$ surface slab unit cell, with 1 adsorbed CO ($\Theta=\frac{1}{4}$). All energies are given in Hartrees, unless stated. Calculations used a $3\times3$ supercell size, using the modified `TZ' basis set. This data was used for Table III, Table IV and Figure 5 of the main text.}
    \label{tab:tdl}
    \begin{tabular}{|l|c|c|c|c|}
    System  & \multicolumn{1}{c|}{HF} & \multicolumn{1}{c|}{$E_{\mathrm{corr}}(\mathcal{T}_{\mathrm{PNO}}=10^{-7})$} & \multicolumn{1}{c|}{$E_{\mathrm{corr}} (\mathcal{T}_{\mathrm{PNO}}=10^{-8})$}&  \multicolumn{1}{c|}{CPS} \\
    \hline
    $[\mathrm{MgOCO}]$          & -2310.10724   & -4.14239 & -4.14407  & -4.14484 \\
    $[\mathrm{MgO,\overline{CO}}]$   & -2197.33047   & -3.77730  & -3.77861  & -3.77921 \\
    $[\mathrm{CO,\overline{MgO}}]$   &  -112.77869   & -0.35809 & -0.35815  & -0.35818 \\
    $[\mathrm{CO}]$             &  -112.77842   & -0.35712  & -0.35714  & -0.35715 \\
    $[\mathrm{CO,\overline{CO}}]$    &  -112.77855   & -0.35685  & -0.35687  & -0.35688 \\ 
    \hline 
    $E_{\mathrm{int}}$      &     0.00204 &-0.00728 & -0.00758  & -0.00772 \\
    $E_{\mathrm{int}} (\mathrm{kJ \; mol^{-1}})$  & 5.37 & -19.11   & -19.90  & -20.28 \\
    \end{tabular}
\end{table*}

\begin{table*}[h]
\centering
    \caption{Hartree--Fock and MP2 correlation energies for each system involved in evaluating $E_{\mathrm{int}}$, employing the $4\cdot4$ surface slab unit cell, with 4 adsorbed COs ($\Theta=\frac{1}{4}$). All energies are given in Hartrees, unless stated. Calculations used a $3\times3$ supercell size, using the modified `TZ' basis set. This data was used for Table III of the main text.}
    \label{tab:tdl}
    \begin{tabular}{|l|c|c|c|c|}
    System  & \multicolumn{1}{c|}{HF} & \multicolumn{1}{c|}{$E_{\mathrm{corr}}(\mathcal{T}_{\mathrm{PNO}}=10^{-7})$} & \multicolumn{1}{c|}{$E_{\mathrm{corr}} (\mathcal{T}_{\mathrm{PNO}}=10^{-8})$}&  \multicolumn{1}{c|}{CPS} \\
    \hline
    $[\mathrm{MgOCO}]$          & -9240.42897   & -16.57109 & -16.57776  & -16.58085 \\
    $[\mathrm{MgO,\overline{CO}}]$   & -8789.32188   & -15.11015  & -15.11593  & -15.11860 \\
    $[\mathrm{CO,\overline{MgO}}]$   &  -451.11475   & -1.43239 & -1.43266  & -1.43278 \\
    $[\mathrm{CO}]$             &  -451.11367   & -1.42849  & -1.42858  & -1.42863 \\
    $[\mathrm{CO,\overline{CO}}]$    &  -112.77855   & -0.35685  & -0.35687  & -0.35688 \\ 
    \hline 
    $E_{\mathrm{int}}$      &     0.00204 &-0.00741 & -0.00757  & -0.00764 \\
    $E_{\mathrm{int}} (\mathrm{kJ \; mol^{-1}})$  & 5.36 & -19.46   & -19.88  & -20.07 \\
    \end{tabular}
\end{table*}

\begin{table*}[h]
\centering
    \caption{Hartree--Fock and MP2 correlation energies for each system involved in evaluating $E_{\mathrm{int}}$, employing the $2\cdot2$ surface slab unit cell, with 2 adsorbed COs ($\Theta=\frac{1}{2}$). All energies are given in Hartrees, unless stated. Calculations used a $3\times3$ supercell size, using the modified `TZ' basis set. This data was used for Table III, Table IV and Figure 5 of the main text.}
    \label{tab:tdl}
    \begin{tabular}{|l|c|c|c|c|}
    System  & \multicolumn{1}{c|}{HF} & \multicolumn{1}{c|}{$E_{\mathrm{corr}}(\mathcal{T}_{\mathrm{PNO}}=10^{-7})$} & \multicolumn{1}{c|}{$E_{\mathrm{corr}} (\mathcal{T}_{\mathrm{PNO}}=10^{-8})$}&  \multicolumn{1}{c|}{CPS} \\
    \hline
    $[\mathrm{MgOCO}]$          & -2422.88403   & -4.51129 & -4.51323  & -4.51412 \\
    $[\mathrm{MgO,\overline{CO}}]$   & -2197.33174   & -3.77881  & -3.78012  & -3.78079 \\
    $[\mathrm{CO,\overline{MgO}}]$   &  -225.55652   & -0.71853 & -0.71871  & -0.71880 \\
    $[\mathrm{CO}]$             &  -225.55602   & -0.71658 & -0.71669  & -0.71674 \\
    $[\mathrm{CO,\overline{CO}}]$    &  -112.77869   & -0.35694  & -0.35696  & -0.35697 \\ 
    \hline 
    $E_{\mathrm{int}}$      &     0.00279 &-0.00833 & -0.00856  & -0.00866 \\
    $E_{\mathrm{int}} (\mathrm{kJ \; mol^{-1}})$  & 7.32 & -21.86   & -22.47  & -22.75 \\
    \end{tabular}
\end{table*}

\begin{table*}[h]
\centering
    \caption{Hartree--Fock and MP2 correlation energies for each system involved in evaluating $E_{\mathrm{int}}$, employing the $4\cdot4$ surface slab unit cell, with 8 adsorbed COs ($\Theta=\frac{1}{2}$). All energies are given in Hartrees, unless stated. Calculations used a $3\times3$ supercell size, using the modified `TZ' basis set. This data was used for Table III of the main text.}
    \label{tab:tdl}
    \begin{tabular}{|l|c|c|c|c|}
    System  & \multicolumn{1}{c|}{HF} & \multicolumn{1}{c|}{$E_{\mathrm{corr}}(\mathcal{T}_{\mathrm{PNO}}=10^{-7})$} & \multicolumn{1}{c|}{$E_{\mathrm{corr}} (\mathcal{T}_{\mathrm{PNO}}=10^{-8})$}&  \multicolumn{1}{c|}{CPS} \\
    \hline
    $[\mathrm{MgOCO}]$          & -9691.53612   & -18.04651 & -18.05435  & -18.05798 \\
    $[\mathrm{MgO,\overline{CO}}]$   & -8789.32695   & -15.11616  & -15.12220  & -15.12500 \\
    $[\mathrm{CO,\overline{MgO}}]$   &  -902.22610   & -2.87418 & -2.87498  & -2.87534 \\
    $[\mathrm{CO}]$             &  -902.22410   & -2.86639  & -2.86685  & -2.86706 \\
    $[\mathrm{CO,\overline{CO}}]$    &  -112.77869   & -0.35694  & -0.35696 & -0.35697 \\ 
    \hline 
    $E_{\mathrm{int}}$      &     0.00279 &-0.00838 & -0.00854  & -0.00861 \\
    $E_{\mathrm{int}} (\mathrm{kJ \; mol^{-1}})$  & 7.32 & -22.01   & -22.42  & -22.62 \\
    \end{tabular}
\end{table*}

\begin{table*}[h]
\centering
    \caption{Hartree--Fock and MP2 correlation energies for each system involved in evaluating $E_{\mathrm{int}}$, employing the $2\cdot2$ surface slab unit cell, with 3 adsorbed COs ($\Theta=\frac{3}{4}$). All energies are given in Hartrees, unless stated. Calculations used a $3\times3$ supercell size, using the modified `TZ' basis set. This data was used for Table IV and Figure 5 of the main text.}
    \label{tab:tdl}
    \begin{tabular}{|l|c|c|c|c|}
    System  & \multicolumn{1}{c|}{HF} & \multicolumn{1}{c|}{$E_{\mathrm{corr}}(\mathcal{T}_{\mathrm{PNO}}=10^{-7})$} & \multicolumn{1}{c|}{$E_{\mathrm{corr}} (\mathcal{T}_{\mathrm{PNO}}=10^{-8})$}&  \multicolumn{1}{c|}{CPS} \\
    \hline
    $[\mathrm{MgOCO}]$          & -2535.64209   & -4.89141 & -4.89375  & -4.89483 \\
    $[\mathrm{MgO,\overline{CO}}]$   & -2197.33217   & -3.77989  &-3.78129  & -3.78194 \\
    $[\mathrm{CO,\overline{MgO}}]$   &  -338.31405   & -1.09127 & -1.09163  & -1.09180 \\
    $[\mathrm{CO}]$             &  -338.31334   & -1.08800 & -1.08825  & -1.08837 \\
    $[\mathrm{CO,\overline{CO}}]$    &  -112.77883   & -0.35723  & -0.35726  & -0.35728 \\ 
    \hline 
    $E_{\mathrm{int}}$      &     0.00910 &-0.01219 & -0.01243  & -0.01254 \\
    $E_{\mathrm{int}} (\mathrm{kJ \; mol^{-1}})$  & 23.88 & -32.00   & -32.64  & -32.93 \\
    \end{tabular}
\end{table*}

\begin{table*}[h]
\centering
    \caption{Hartree--Fock and MP2 correlation energies for each system involved in evaluating $E_{\mathrm{int}}$, employing the $2\cdot2$ surface slab unit cell, with 4 adsorbed COs ($\Theta=1$). All energies are given in Hartrees, unless stated. Calculations used a $3\times3$ supercell size, using the modified `TZ' basis set. This data was used for Table IV and Figure 5 of the main text.}
    \label{tab:tdl}
    \begin{tabular}{|l|c|c|c|c|}
    System  & \multicolumn{1}{c|}{HF} & \multicolumn{1}{c|}{$E_{\mathrm{corr}}(\mathcal{T}_{\mathrm{PNO}}=10^{-7})$} & \multicolumn{1}{c|}{$E_{\mathrm{corr}} (\mathcal{T}_{\mathrm{PNO}}=10^{-8})$}&  \multicolumn{1}{c|}{CPS} \\
    \hline
    $[\mathrm{MgOCO}]$          & -2648.39932   & -5.27365 & -5.27588  & -5.27691\\
    $[\mathrm{MgO,\overline{CO}}]$   & -2197.33265   & -3.78105  &-3.78247  & -3.78312 \\
    $[\mathrm{CO,\overline{MgO}}]$   &  -451.071166   & -1.46611 & -1.46677  & -1.46707 \\
    $[\mathrm{CO}]$             &  -451.07015   & -1.46138 & -1.46183  & -1.46203 \\
    $[\mathrm{CO,\overline{CO}}]$    &  -112.77900   & -0.35754  & -0.35759  & -0.35761 \\ 
    \hline 
    $E_{\mathrm{int}}$      &     0.01258 &-0.01442 & -0.01453 & -0.01458 \\
    $E_{\mathrm{int}} (\mathrm{kJ \; mol^{-1}})$  & 33.04 & -37.87     & -38.15  & -38.28 \\
    \end{tabular}
\end{table*}

\clearpage
\newpage
\printbibliography[title={References}]